\documentclass{tlp}

\usepackage{amssymb}
\usepackage{aopmath}
\usepackage{psfrag}
\usepackage{graphics}

\newcommand{\tran}[1]{\stackrel{#1}{\rightarrow}}
\newcommand{\ttran}[1]{\stackrel{#1}{\mapsto}}

\newcommand{\Tran}[1]{\stackrel{#1}{\Rightarrow}}

\newcommand{\Nseto}{\mathrm{I\!N}}
\newcommand{\Natomega}{\mathrm{I\!N}_\omega}

\newcommand{\sm}{\sqsubseteq_{\mathit{sm}}}
\newcommand{\sme}{=_{\mathit{sm}}}
\newcommand{\tr}{\sqsubseteq_{\mathit{tr}}}
\newcommand{\tre}{=_{\mathit{tr}}}
\newcommand{\wsim}{\approx}

\newcommand{\Ex}[1]{\langle #1 \rangle}
\newcommand{\Al}[1]{[#1]}
\newcommand{\wEx}[1]{\langle\!\langle #1 \rangle\!\rangle}

\newcommand{\kw}[1]{\textbf{#1}}

\newcommand{\ttt}{\mathtt{tt}}
\newcommand{\fff}{\mathtt{ff}}

\newcommand{\G}{\mathcal{G}}
\newcommand{\F}{\mathcal{F}}
\newcommand{\A}{\mathcal{A}}
\newcommand{\B}{\mathcal{B}}
\newcommand{\C}{\mathcal{C}}
\newcommand{\E}{\mathcal{E}}
\newcommand{\T}{\mathcal{T}}
\newcommand{\M}{\mathcal{M}}
\newcommand{\N}{\mathcal{N}}

\renewcommand{\P}{\mathcal{P}}
\renewcommand{\S}{\mathcal{S}}
\newcommand{\calO}{\mathcal{O}}

\newcommand{\Act}{\mathit{Act}}

\newcommand{\Control}{\mathit{Control}}
\newcommand{\Stack}{\mathit{Stack}}

\newcommand{\Congr}{\mathit{Congr}}
\newcommand{\Reach}{\mathit{Reach}}
\newcommand{\Gen}{\mathit{Gen}}
\newcommand{\Bis}{\mathit{Bis}}

\newcommand{\norm}{\mathit{norm}}
\newcommand{\length}{\mathit{length}}
\newcommand{\PDA}{\textbf{PDA}}
\newcommand{\PPDA}{\textbf{PPDA}}
\newcommand{\BPA}{\textbf{BPA}}
\newcommand{\BPP}{\textbf{BPP}}
\newcommand{\PA}{\textbf{PA}}
\newcommand{\nBPA}{\textbf{nBPA}}
\newcommand{\nBPP}{\textbf{nBPP}}
\newcommand{\nPA}{\textbf{nPA}}
\newcommand{\PN}{\textbf{PN}}
\newcommand{\OCN}{\textbf{OC-N}}
\newcommand{\OCA}{\textbf{OC-A}}
\newcommand{\FS}{\textbf{FS}}

\newcommand{\PSPACE}{\mbox{\textbf{PSPACE}}}
\newcommand{\EXPTIME}{\mbox{\textbf{EXPTIME}}}
\newcommand{\PTIME}{\mbox{\textbf{P}}}
\newcommand{\NP}{\mbox{\textbf{NP}}}
\newcommand{\coNP}{\mbox{\textbf{coNP}}}
\newcommand{\DP}{\mbox{\textbf{DP}}}
\newcommand{\EXPSPACE}{\mbox{\textbf{EXPSPACE}}}

\newcommand{\dist}{\mathit{dist}}
\newcommand{\mynorm}{\textrm{\sc norm}}

\newtheorem{definition}{Definition}
\newtheorem{example}{Example}

\submitted{28 November 2003}
\revised{20 September 2004}
\accepted{10 May 2005} 

\begin{document}

\title[Equivalence-Checking on Infinite-State Systems]{Equivalence-Checking 
    on Infinite-State Systems: Techniques and Results
%PJ
% ja chci u sebe jen grant, ktery uvadim (Duzi, informacni
% spolecnost; u neho zrejme
% budu participovat jen 07 - 12/2004)
% \thanks{Supported 
%    by the Grant Agency of Czech Republic, grant No.\ 201/03/1161.}
}

\author[A. Ku\v{c}era and P. Jan\v{c}ar]%
    {ANTON\'{I}N KU\v{C}ERA\thanks{Supported by the
     research center Institute for Theoretical Computer Science (ITI),
     project No.~1M0021620808.}\\
     Faculty of Informatics, Masaryk University, Botanick\'a 68a,\\ 
     CZ-602~00 Brno, Czech Republic.\\
     \email{tony@fi.muni.cz}
     \and
     PETR JAN\v{C}AR\thanks{Supported by the grant 
     No.\ 1ET101940420 of the Czech national program 
     ``Information society''.}\\
     Dept.\ of Computer Science, FEI, Technical University of Ostrava,
     17.\ listopadu 15,\\ 
     CZ-708~33 Ostrava, Czech Republic.\\
     \email{Petr.Jancar@vsb.cz}} 

\maketitle

\begin{abstract}
  The paper presents a selection of recently developed and/or used
  techniques for equivalence-checking on infinite-state systems, and
  an up-to-date overview of existing results (as of September 2004).
\end{abstract}

\section{Introduction} 
\label{sec-intro}

A \emph{reactive system} is a system which continuously interacts with
its environment and whose behavior is strongly influenced by this
interaction. Reactive systems usually consist of several asynchronous
(but communicating) processes which run in parallel.  This asynchrony,
together with unpredictable actions of the environment, contribute to
a high degree of non-determinism.  Another characteristic feature is
divergence; a reactive system is often supposed to run forever, though
its processes can be dynamically created and terminated. Since
reactive systems control potentially dangerous devices like power
plants, airports, weapon systems, etc., there is a strong need for
rigorous methods which allow to \emph{prove} correctness (or at least
safety) of such systems.

Two popular approaches to formal verification of reactive systems are
\emph{model-checking} and \emph{equivalence-checking}. In the
model-checking approach, desired properties of the verified
implementation are defined as a formula of a suitable modal logic, and
then it is shown that (a formal model of) the implementation satisfies
the formula.  In the equivalence-checking approach, one constructs a
formal model of the intended behavior of the verified system (called
\emph{specification}) and then it is shown that the
implementation is equivalent to the specification.

A principal difficulty of automated formal verification is that
reactive systems tend to have a very large state space. There are
various strategies for tackling this problem. For example, the
technique of \emph{symbolic model-checking} introduced in
\cite{BCMDH:symbolic-IC} uses a symbolic state-space representation
based on OBDD's (ordered binary decision diagrams). This method was
successfully used for formal verification of hardware circuits.
\emph{Partial-order reduction} (as described, e.g., in
\cite{CGP:book}) enables a practical verification of concurrent
software based on model-checking with the logic LTL. Though these
methods handle systems with large state spaces, they are still limited
to finite-state systems. However, many systems are (or should be seen
as) unbounded, i.e., having a potentially infinite state space.  For
example, unbounded data types such as counters, stacks, channels, or
queues, require an infinite number of states. Parametrized systems
(e.g., $N$ philosophers, $N/M$ readers/writers, etc.) should also be
seen as infinite-state if we want to show their correctness for every
choice of parameters. Another example are systems with a dynamically
evolving structure (e.g., mobile networks).

Model-checking and equivalence-checking on infinite-state systems
is a popular research field which has been attracting attention 
for almost two decades. Consequently, the collection of achieved results
is large and diverse today. There have been several surveys presenting 
various subfields of this research area, like 
\cite{Moller:infinite_results,Esparza:ModelChecking-AI,%
  JM:Techniques,Bouajjani:verification,Srba:Roadmap}, including a
major Handbook chapter~\cite{BCMS:Infinite-Structures-HPA}.  
This paper is intended as a contribution to the collection of surveys, 
and its aim is twofold. First, it presents a selection of some recently
developed techniques for equivalence-checking on infinite-state
systems which have not yet been fully covered in the existing surveys. The
emphasis is on explaining the core of underlying principles rather
than presenting full proofs of particular results.  Second, the paper
gives an up-to-date overview of existing results for
equivalence-checking on infinite-state systems (as of September 2004).

The style of presentation adopted in this paper reflects the 
authors' intention to explain ``proof techniques'' rather than
particular proofs. Ideally, this would be achieved by first formulating 
a given technique ``abstractly'', and then showing how it applies
in concrete situations. In most cases, we provide a detailed 
explanation just for the ``abstract'' part,
and then indicate how and where the principle can be applied without
going much into details (just pointing to the relevant literature). 
When we feel that the abstract formulation is too vague, the
functionality is demonstrated on concrete examples.

The paper is organized as follows. Section~\ref{sec-defs} contains
basic definitions. Section~\ref{sec-techniques} is devoted to the
presentation of selected proof techniques. In particular,
Section~\ref{sec-sim-bisim} presents general results about the
relationship between simulation preorder/equivalence and bisimulation
equivalence.
  Subsection~\ref{sec-bisim-to-sim} starts by a simple observation 
  about a specific power of the defender in simulation games. 
  This observation is then used in a general reduction scheme
  which allows to (efficiently) reduce bisimilarity problems to
  their simulation counterparts.
  In Subsection~\ref{sec-sim-to-bisim} it is shown that there is also
  a generic ``reduction'' of the simulation equivalence problem to
  the bisimilarity problem. Although this ``reduction'' is rarely effective
  (due to fundamental reasons), it reveals a simple and generic relationship
  between simulation equivalence and bisimilarity.

Section~\ref{sec-dec-upper} is devoted to selected techniques which
  have recently been used to establish new decidability results and upper
  complexity bounds for equivalence-checking problems.
  In Subsection~\ref{sec-bases}, the technique of \emph{bisimulation
  bases} is recalled (in a somewhat ``abstracted'' form) and then it is shown
  how this technique applies to checking weak bisimilarity between
  infinite and finite-state systems.
  In Subsection~\ref{sec-equiv-FS}, the problem of effective
  constructibility of \emph{characteristic formulae} which express the 
  equivalence with a given finite-state system is examined. First, well-known
  results about the constructibility of characteristic formulae
  in the modal $\mu$-calculus are recalled. Then, it is shown
  how to construct characteristic formulae w.r.t.\ (strong and weak)
  bisimilarity in the simpler logic EF. 
  In Subsection~\ref{sec-DD-functions}, the so-called
  \emph{DD-functions} are presented. This is a recently discovered ``tool'' 
  used for several decidability and complexity results.

In Section~\ref{sec-undec-lower} we discuss techniques for 
  undecidability and lower complexity bounds. A common principle which
  is used in almost all undecidability and hardness proofs for bisimilarity-
  and simulation-checking problems is the ability of the defender to
  ``force'' the attacker to perform a specific transition. The variant
  for simulation-checking is, in fact, discussed already
  in Subsection~\ref{sec-bisim-to-sim}; a similar
  principle exists also for bisimilarity. Since the abstract formulation
  of the two techniques does not say much about their applicability, 
  we demonstrate them on selected examples.

Section~\ref{sec-results} contains an up-to-date overview of existing
results.

\section{Basic Definitions}
\label{sec-defs}

The set of all non-negative integers $0,1,2,\dots$ is denoted by $\Nseto$.
The symbol $\omega$ is used to denote an infinite amount.

The first step of formal verification is to create
a \emph{formal model} of the verified system. The low-level semantics
of such a model is given by its associated 
\emph{transition system}; in our framework we assume
that transitions (between states) are labelled by actions taken
from a finite set.  

\begin{definition}
  A \emph{transition system} is a triple $\T = (S,\Act,\tran{})$ where
  $S$ is a set of \emph{states}, $\Act$ is a finite set of \emph{actions},
  and ${\tran{}} \subseteq S {\times} \Act {\times} S$ is a \emph{transition
  relation}.
\end{definition}
Processes are formally understood as states in transition
systems;
from now on we do not distinguish between ``states'' and ``processes''.
The dynamics of processes, i.e., possible computational steps, are
defined by the transition relation. We write $s \tran{a} t$ instead
of $(s,a,t) \in {\tran{}}$, and say that $t$ is an $a$-successor of $s$.
This notation is extended to finite strings
over $\Act$ in the natural way. A state $t$ is \emph{reachable} from
a state $s$, written $s \tran{}^* t$, if there is $w \in \Act^*$ such
that $s \tran{w} t$. A transition system is \emph{image-finite}
if each state has only finitely many $a$-successors for every $a \in \Act$.
The \emph{branching degree} of a transition system $\T$, 
denoted $d(\T)$, is the least
$k \in \Nseto$ such that every state of $\T$ has at most $k$ successors
(if there is no such $k$ then $d(\T) = \infty$).

\subsection{Behavioral Equivalences}

The notion of process equivalence can be formalized in many different ways
\cite{Glabbeek:hierarchy-PA-handbook,Glabbeek:hierarchy-weak}.  
A straightforward idea is to
employ the classical notion of language equivalence from automata
theory (here we consider all states as accepting):

\begin{definition}
Let $\T = (S,\Act,\tran{})$ be a transition system, $s \in S$.  
We say that $w \in \Act^*$ is a \emph{trace} of
$s$ iff $s \tran{w} s'$ for some $s'$. Let $\mathit{tr}(s)$ be the set
of all traces of $s$. We write $s \tr t$ iff $\mathit{tr}(s) \subseteq
\mathit{tr}(t)$.  Moreover, we say that $s$ and $t$ are \emph{trace
equivalent}, written $s \tre t$, iff $\mathit{tr}(s) = \mathit{tr}(t)$.
\end{definition}

\begin{figure}
\centering
\psfrag{s}[c][c]{\scriptsize\mbox{$s$}}
\psfrag{t}[c][c]{\scriptsize\mbox{$t$}}
\psfrag{u}[c][c]{\scriptsize\mbox{$u$}}
\psfrag{la}[l][l]{\scriptsize\mbox{$a$}}
\psfrag{ra}[r][r]{\scriptsize\mbox{$a$}}
\psfrag{b}[r][r]{\scriptsize\mbox{$b$}}
\psfrag{c}[l][l]{\scriptsize\mbox{$c$}}
\includegraphics{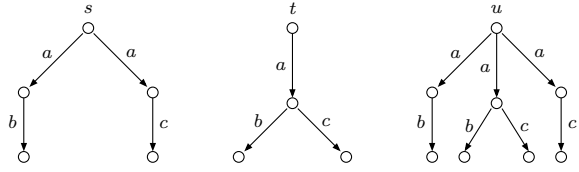}
\caption{Processes $s$, $t$, and $u$.}
\label{fig-exa-equiv}
\end{figure}

In concurrency theory, trace equivalence is usually considered as being
too coarse. For example, the processes $s$ and $t$ of 
Fig.~\ref{fig-exa-equiv} are trace equivalent but their behavior
is different---$s$ can do either $b$ or $c$ (but not both) 
after performing $a$, while $t$ can always choose between $b$ and $c$
after $a$.  A finer level of ``semantical sameness'' of two processes can be
defined by formalizing the ability of one process to ``mimic'' (or 
\emph{simulate}) computational steps of another process.

\begin{definition}
Let $\T = (S,\Act,\tran{})$ be a transition system, $s,t \in S$. 
A binary relation $R$ over $S$ is a \emph{simulation}
iff whenever $(s,t) \in R$ then for every $a \in \Act$
\[
  \mbox{if } s \tran{a} s' \mbox{ then } t \tran{a} t' 
  \mbox{ for some } t' \mbox{ such that } (s',t') \in R\mbox{.}
\]
A process $s$ is \emph{simulated} by a process $t$, written $s \sm t$,
iff there is a simulation $R$ such that $(s,t) \in R$. 
Note that the relation $\sm$ is a preorder.
We say that $s$ and
$t$ are \emph{simulation equivalent}, written $s \sme t$, iff $s \sm t$
and $t \sm s$.
\end{definition}
For example, for processes of Fig.~\ref{fig-exa-equiv} we have that
$s \sm t$, $t \not\sm s$, and $t \sme u$.

Simulation preorder and equivalence can also be defined in terms of
games \cite{Stirling:book,Thomas:games}. 
Imagine there are two tokens put on states $s$ and $t$. Two
players, the \emph{attacker} and the \emph{defender}, start to play 
a \emph{simulation game} which
consists of (possibly infinite) sequence of \emph{rounds}, where each round
is performed as follows:
\begin{enumerate}
\item the attacker takes the first token (the one which was put on $s$ 
   originally) and moves it along an arbitrary transition labeled by some
   $a \in \Act$;
\item the defender has to respond by moving the other token along 
   some transition with the same label $a$.
\end{enumerate}
One player \emph{wins} if the other player cannot
move. Moreover, the defender wins every infinite play.
It is easy to see that $s \sm t$ iff the defender has 
a universal winning strategy. Simulation equivalence can be 
understood similarly; we simply allow the attacker to choose his token
at the beginning of the first round.

The finest (and probably the most important) behavioral equivalence
we consider is \emph{bisimulation equivalence} 
\cite{Park:bisimulation,Milner:book}. 
\begin{definition}
Let $\T = (S,\Act,\tran{})$ be a transition system, $s,t \in S$. 
A binary relation $R$ over $S$ is a \emph{bisimulation}
iff whenever $(s,t) \in R$ then for every $a \in \Act$
\begin{itemize}
\item if $s \tran{a} s'$ then $t \tran{a} t'$ for some $t'$ such that
   $(s',t') \in R$,
\item if $t \tran{a} t'$ then $s \tran{a} s'$ for some $s'$ such that
   $(s',t') \in R$.
\end{itemize}
Processes $s,t$ are \emph{bisimulation equivalent} (or \emph{bisimilar}),   
written $s \sim t$, iff there is a bisimulation $R$ such that $(s,t) \in R$.
\end{definition}
A \emph{bisimulation game} is defined in the same way as
the simulation game. 
The only difference is that the attacker can 
\emph{choose} his token at the beginning of \emph{every} round
(the defender has to respond with the other token). Again we have that 
$s \sim t$ iff the defender has a universal 
winning strategy in the bisimulation game initiated in $s,t$. For
example, one can check that the processes $s,t,u$ of Fig.~\ref{fig-exa-equiv}
are pairwise non-bisimilar. 

Internal computational steps which are not directly observable are by
convention denoted by a special action $\tau$.  The notion of
\emph{weak bisimilarity} \cite{Milner:book} allows to ``ignore'' the
internal steps to some extent.
\begin{definition}
\label{def-weak-bisimulation}
Let $\T = (S,\Act,\tran{})$ be a transition system.  The
\emph{extended transition relation} ${\Tran{}} \subseteq S {\times}
\Act {\times} S$ is defined as follows: $s \Tran{a} t$ iff one of the
two conditions~holds:
\begin{itemize}
\item $a \neq \tau$ and there are $s',s'' \in S$, $i,j \in \Nseto$ such that 
  $s \tran{\tau^i} s' \tran{a} s'' \tran{\tau^j} t$.
\item $a = \tau$ and there is $i \in \Nseto$ such that $s \tran{\tau^i} t$.
\end{itemize}
Here $s \tran{\tau^0} s'$ iff $s = s'$.
In particular, this means that $s \Tran{\tau} s$ for every $s \in S$.
A binary relation $R$ over $S$ is a \emph{weak bisimulation}
iff whenever $(s,t) \in R$ then for every $a \in \Act$
\begin{itemize}
\item if $s \Tran{a} s'$ then $t \Tran{a} t'$ for some $t'$ such that
   $(s',t') \in R$,
\item if $t \Tran{a} t'$ then $s \Tran{a} s'$ for some $s'$ such that
   $(s',t') \in R$.
\end{itemize}
Processes $s,t$ are \emph{weakly bisimulation equivalent} (or
\emph{weakly bisimilar}), written $s \wsim t$, iff there is a weak
bisimulation $R$ such that $(s,t) \in R$.
\end{definition}
A \emph{weak bisimulation game} is defined in the same way as
the bisimulation game, but both players now use the extended transitions.

We say that processes $s$ and $t$ are \emph{bisimilar up to $i \in \Nseto$},
written $s \sim_i t$, if the defender has a winning
strategy for the first $i$ rounds of the bisimulation game initiated in
$s$ and $t$.  It is easy to see that $\sim_i$ is an equivalence
relation and that $\sim_{i+1}$ refines $\sim_i$ for every $i \in
\Nseto$. Also note that $s \sim_0 t$ for all processes $s,t$. An important
observation, taken from \cite{BBK:consistency-TCS}, is
\begin{theorem}
\label{thm-finitely-branching}
  Let $\T = (S,\Act,\tran{})$ be a transition system and let $s,t$ be 
  processes of $\T$ such that each state $t'$ reachable from $t$ has
  only finitely many $a$-successors for every $a \in \Act$ (note that
  there is no assumption about the process $s$). Then
  $s \sim t$ iff $s \sim_i t$ for every $i \in \Nseto$.
\end{theorem}
\begin{proof}
  The ``$\Longrightarrow$'' is obvious. For the other direction, one
  can check that the relation 
  $R = \{(s',t') \mid (\forall i \in \Nseto: s' \sim_i t')
  \,\wedge\, t \tran{}^*t' \}$ is a bisimulation: Since 
  $t'$ has finitely many $a$-successors, for each 
  $s' \tran{a} s''$ there must be some $t' \tran{a} t''$ such that 
  $\forall i \in \Nseto: s'' \sim_i t''$. Now consider 
  a move $t' \tran{a} t''$. Obviously, for each $i \in \Nseto$ there is
  $s' \tran{a} s_i$ such that $s_i \sim_i t''$. 
  Each of the $s' \tran{a} s_i$ moves must be matched by some 
  transition of $t'$. Since $t'$ has only finitely many $a$-successors, 
  there is a transition $t' \tran{a} t'''$ which was used infinitely
  many times. That is, there is an infinite sequence $s_{i_1}, s_{i_2},\dots$  
  such that for each $s_{i_j}$ we have
  $\forall i\in \Nseto: s_{i_j} \sim_i t'''$. This
  means $\forall i\in \Nseto: t''' \sim_i t''$, and hence
  for every $s_{i_j}$ we have 
  $\forall i\in \Nseto: s_{i_j} \sim_i t''$.
\end{proof}
\emph{Weak bisimilarity up to $i \in \Nseto$}, denoted $\wsim_i$, is
defined in the same way (we use the weak bisimulation game).
The aforementioned observations about $\sim_i$ are valid
also for $\wsim_i$ (incl.\ Theorem~\ref{thm-finitely-branching}
where the $a$-successors are considered w.r.t.\ $\Tran{a}$).

Behavioral equivalences can also be used to relate processes of
\emph{different} transition systems. Formally, we can consider
two transition systems to be a \emph{single} one by taking their
disjoint union (the labeling of transitions is preserved).

The relationship among the introduced equivalences is given by ${\tre}
\supset {\sme} \supset {\sim}$. Weak bisimilarity properly subsumes
$\sim$ and is incomparable with $\tre$ and $\sme$.  (We do not
consider weak versions of trace equivalence and simulation equivalence
in this paper.)  There are also other behavioral preorders and
equivalences studied within the framework of concurrency theory. It
seems, however, that trace, simulation, and especially (weak)
bisimulation equivalence are of special importance as their
accompanying theories are developed very intensively.  Moreover, each
equivalence in the linear/branching time spectrum of
\cite{Glabbeek:hierarchy-PA-handbook} can be classified either as
trace-like or as simulation-like. This means that $\tre$, $\sme$, and
$\sim$ are good representatives for the whole spectrum; techniques and
results achieved for these equivalences usually extend to others.

\subsection{Formal Models of Infinite-State Systems}
\label{sec-models}

In this section we formally introduce some of the 
studied models of infinite-state systems.  At a certain level of
abstraction, most of them can be seen as various types of term
rewriting systems. The structure of terms represents both control and
data of the system, and the individual rewriting steps model atomic
computational steps.  

We start with the definition of a general process rewrite system (PRS)
\cite{Mayr-PRS-IC}.  Then, we define various subclasses of PRS by imposing 
certain restrictions on the introduced formalism.

We assume a countable infinite set $\C$ of 
(process) \emph{constants}. The abstract syntax
of general \emph{process expressions} is given by
\[
  E \quad ::= \quad X \quad | \quad \varepsilon \quad | 
              \quad E.E \quad | \quad E\|E
\]
where the (meta)variable $X$ ranges over 
$\C$ and $\varepsilon$ denotes the empty expression. 
Intuitively, ``$.$'' corresponds to sequencing, while ``$\|$'' models a
simple form of parallelism.  From now on we do not distinguish between
expressions related by the \emph{structural congruence}, which is the
smallest congruence over $\E$ satisfying the following laws: ``$.$'' and
``$\|$'' are associative, $\varepsilon$ is the unit for both operators,
and ``$\|$'' is also commutative.  

The set of all process expressions is denoted by $\E$.  The sets of
\emph{sequential} and \emph{parallel} expressions, denoted $\S$ and
$\P$, are formed by all process expressions which do not contain any
``$\|$'' and ``$.$'', respectively. Observe that parallel expressions
can also be seen as \emph{multisets} of constants.  Given
$\C'\subseteq \C$, we use $\S(\C')$, $\P(\C')$, and $\E(\C')$ to
denote the set of all sequential expressions, parallel expressions,
and general expressions, respectively, which contain only the
constants from $\C'$.

We also assume a countable infinite set $\A$
of \emph{actions}, ranged over by $a,b,c,\dots$\,.
A \emph{process rewrite system (PRS)} is a \emph{finite} subset $\Delta$ of
$\E \times \A \times \E$. Elements of $\Delta$ are called \emph{rules}
(a rule $(\alpha,a,\beta)$ is usually 
written $\alpha \tran{a} \beta$).
Given a PRS $\Delta$, we use $\C(\Delta)$ to denote the set of all constants
appearing in the rules of $\Delta$. We also use $\S(\Delta)$, 
$\P(\Delta)$, and $\E(\Delta)$ to denote  
$\S(\C(\Delta))$, 
$\P(\C(\Delta))$, and $\E(\C(\Delta))$ respectively.
Moreover, $\A(\Delta)$ denotes the set of actions
which are used in the rules of $\Delta$.

Each PRS $\Delta$ determines a unique transition system $\T_\Delta$ where
$\E(\Delta)$ is the set of states, $\A(\Delta)$ is the set of actions,
and the transition relation is determined by the following inference rules
(which should be understood modulo the structural congruence over
expressions introduced above):
\[
\newcommand{\mysize}{\small}
\frac{\mysize\mbox{$(E \tran{a} F) \in \Delta$}}%
     {\mysize\mbox{$E \tran{a} F$}}
\hspace*{2em} 
\frac{\mysize\mbox{$E \tran{a} F$}}%
     {\mysize\mbox{$E.G \tran{a} F.G$}}
\hspace*{2em}
\frac{\mysize\mbox{$E \tran{a} F$}}%
     {\mysize\mbox{$E\|G \tran{a} F\|G$}}
\]
Various subclasses of PRS can be obtained by imposing certain
restrictions on the form of the rules. Such a restriction is formally
specified by a pair $(A,B)$, where $A$ and $B$ are the subsets of
expressions which can appear at the left-hand side and the right-hand side
of rules, respectively. It has been argued in \cite{Mayr-PRS-IC}
that ``reasonable'' restrictions should satisfy $A \subseteq B$. 
Moreover, if $\Delta$ is an $(A,B)$-restricted 
PRS, then the set of states of $\T_\Delta$ is restricted to 
$B \cap \E(\Delta)$. Some of the most important subclasses of PRS 
are listed below.
\begin{itemize}
\item \emph{Finite state (FS) systems}. These are $(\C,\C)$-restricted
   PRS which correspond to ``ordinary'' nondeterministic finite automata; 
   the only difference is that there are no initial/final states. 
\item \emph{BPA systems}. The restriction is $(\C,\S)$. This model
   corresponds to the BPA (Basic Process Algebra) 
   fragment of ACP \cite{BW:book}.
%PJ co je vlastne ACP ? take uvest slovy ?
%recenzent to chce u vsech zkratek
\item \emph{BPP systems}. The restriction is $(\C,\P)$.
   BPP (Basic Parallel Processes) first 
   appeared in the work 
   \cite{Christensen:PhD}.
\item \emph{PA systems}. The restriction is $(\C,\E)$. 
   PA (Process Algebra) systems subsume both BPA and BPP systems 
   and correspond 
   to another natural fragment of ACP \cite{BW:book}.
\item \emph{PDA systems}. The restriction is $(\S,\S)$. 
   It has been shown in \cite{Caucal:prefix-rewr-TCS} that every 
   PDA system $\Delta$ can 
   be efficiently transformed to a ``normal form'' $\Delta'$ 
   where 
   \begin{itemize}
   \item the set $\C(\Delta')$ can be partitioned into two
     disjoint subsets $\Control(\Delta')$ and $\Stack(\Delta')$;
   \item the rules are of the form $p \cdot X \tran{a} q \cdot \beta$
     where $p,q \in \Control(\Delta')$, $X \in \Stack(\Delta')$, and 
     $\beta \in \S(\Stack(\Delta'))$;
   \item the set of states of $\T_{\Delta'}$ is restricted 
     to those elements of $\S(\Delta')$ which are of the form
     $p \cdot \alpha$ where $p \in \Control(\Delta')$ and
     $\alpha \in \S(\Stack(\Delta'))$.
   \end{itemize}
   Hence, PDA systems correspond to pushdown automata \cite{HU:book}.
   Consistently with the standard notation, we write $p \alpha$ instead
   of $p \cdot \alpha$.
   Observe that BPA can be also seen as PDA with just one control state.
\item \emph{PN systems}. The restriction is $(\P,\P)$.
   PN systems correspond to the well-known model of Petri nets.
   Here the elements of 
   $\C(\Delta)$ are referred to as \emph{places} and the states 
   of $\T_\Delta$ (i.e., multisets of places) as \emph{markings}. 
   In the rest of this paper we use the standard graphical representation of
   Petri nets to define PN systems---places are depicted as circles, 
   and for every rule $X_1\| \dots \| X_n \tran{a} Y_1\| \dots \|Y_n$
   we draw a new 
   square labeled by ``$a$''. The square is connected to every $X_i$ 
   by an arrow pointing to the square, and to every $Y_j$ by an arrow 
   pointing to $Y_j$. For example, the middle part of 
   Fig.~\ref{fig:pj-undec} represents the rule
   $Q_i\|C_j \tran{dec} Q_l$, the right-hand part represents the
 rules $Q_i \tran{zer} Q_k$, 
   $Q_i\|C_j \tran{zer} Q'_k\|C_j$ etc.
\item \emph{PPDA systems}. This is a subclass of PN known as 
   ``Parallel PushDown Automata'' \cite{Moller:infinite_results}. 
   A system $\Delta$ is PPDA
   if the set $\C(\Delta)$ can be partitioned into two
   disjoint subsets $\Control(\Delta)$ and $\Stack(\Delta)$ so that
   every rule of $\Delta$ is of the form $p \| X \tran{a} q \| \beta$
     where $p,q \in \Control(\Delta)$, $X \in \Stack(\Delta)$, and 
     $\beta \in \P(\Stack(\Delta))$. 

   For a PPDA system $\Delta$, the set of states of $\T_\Delta$ is
   restricted to those elements of $\P(\Delta)$ which are of the form
     $p \| \alpha$ where $p \in \Control(\Delta)$ and
     $\alpha \in \P(\Stack(\Delta))$. Usually we write $p \alpha$ instead
   of $p \| \alpha$.
\item \emph{OC-A systems}. These are PDA systems in normal form such that
   $\Stack(\Delta) = \{I,Z\}$ and all transitions are of 
   the form $pZ \tran{a} qI^iZ$ or $rI \tran{a} sI^j$, where 
   $i,j \geq 0$. Here $I^i$ denotes
   the sequential composition of $i$ copies of the symbol $I$. 
   The set of states of $\T_\Delta$ is restricted to 
   $Q {\times} \{I^iZ \mid i \geq 0\}$. Hence, OC-A systems are one-counter
   automata where the counter ranges over nonnegative values. The counter
   can be incremented, decremented (if positive), and tested for zero.
\item \emph{OC-N systems}. These are OC-A
   systems which in addition satisfy the following condition:
   if $pZ \tran{a} qI^iZ$ is a rule of $\Delta$, then
   also $pI \tran{a} qI^iI$ is a rule of $\Delta$. In other words,
   there are no ``zero-specific'' transitions which could be used to
   test the counter for zero. OC-N systems are equivalent to Petri nets
   with at most one unbounded place. 
\end{itemize}
Let \emph{C} be one of the just defined subclasses of PRS.
A \emph{C-process} is a state in $\T_\Delta$ where $\Delta$ is a
member of \emph{C}. 
The class of all \emph{C}-processes is denoted $\textbf{C}$.
Important subclasses of BPA, BPP, and PA systems can be obtained by an
extra condition of \emph{normedness}. A BPA, BPP, or PA system 
$\Delta$ is \emph{normed} if for every $X \in \C(\Delta)$ we have
$X \tran{}^* \varepsilon$. Hence, a system is normed if each of its
processes can terminate via a finite number of transitions.
The normed subclasses of BPA, BPP, and PA are denoted
by $\nBPA$, $\nBPP$, and $\nPA$, respectively. 

Let $\leq$ be an ordering over process classes 
defined by $\textbf{C}_1 \leq \textbf{C}_2$ 
iff for every $C_1$-process there is a bisimilar $C_2$-process.
The relationship among the introduced subclasses of processes 
(w.r.t.\ $\leq$) is shown in the following figure (we
refer to \cite{Moller:infinite_results} for results about 
expressiveness).
\begin{center}
{\scriptsize
\psfrag{FS}[c][c]{$\FS$}
\psfrag{BPP}[c][c]{$\BPP$}
\psfrag{BPA}[c][c]{$\BPA$}
\psfrag{OCN}[c][c]{$\OCN$}
\psfrag{OCA}[c][c]{$\OCA$}
\psfrag{PA}[c][c]{$\PA$}
\psfrag{PN}[c][c]{$\PN$}
\psfrag{PDA}[c][c]{$\PDA$}
\psfrag{PPDA}[c][c]{$\PPDA$}
\includegraphics{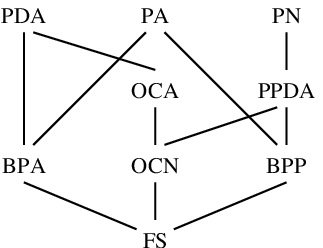}}
\end{center}
Let $\simeq$ be a relation over processes. The problem of deciding
$\simeq$ between processes of process classes $\textbf{A}$ and $\textbf{B}$
is denoted $\textbf{A} \simeq \textbf{B}$. For example, the problem
of deciding bisimilarity between BPA and BPP processes is denoted
$\BPA \sim \BPP$, and the problem of deciding simulation preorder
between PA a FS processes is denoted $\PA \sm \FS$.

\section{Some Recent Techniques and Results}
\label{sec-techniques}

In this section we explain some techniques which
have recently been used to establish new
decidability/complexity results for
equivalence-checking on infinite-state systems.  The material is
divided into three (sub)sections. In Section~\ref{sec-sim-bisim} we
explore the relationship between bisimilarity and simulation
equivalence. Section~\ref{sec-dec-upper} sketches some techniques for
decidability and upper complexity bounds.
Section~\ref{sec-undec-lower} deals with techniques for undecidability and
lower complexity bounds. 

The generality and versatility of proof techniques is of course hard
to measure.  In the context of equivalence-checking on
infinite-state systems, one good indication of a wider applicability
of a given technique is a possibility to formulate its underlying
principle in terms of transition systems (then we can say that the
technique is ``implemented'' in a given syntax). However, such a
formulation is not always possible despite a clear feeling that many
proofs are just ``instances'' of the same idea. Here, we have to rely on
an informal explanation and present an example which uses the
technique in its simple and ``clean'' form.

\subsection{The Relationship Between Simulation and Bisimulation}
\label{sec-sim-bisim}

Since formal definitions of simulation and bisimulation are quite
similar, a natural question is whether the decidability/complexity results
achieved for one of the equivalences carry over to the other one.
In this section we examine the question in greater detail. 

\subsubsection{Reducing Bisimilarity to Simulation Preorder/Equivalence.}
\label{sec-bisim-to-sim}

According to the known decidability/complexity results for simulation and
bisimilarity
(which will be presented in Section~\ref{sec-results}), the problems 
$\textbf{A} \sm \textbf{B}$ and $\textbf{A} \sme \textbf{B}$ are
computationally harder than the problem $\textbf{A} \sim \textbf{B}$
for all major process classes $\textbf{A}$ and $\textbf{B}$.
The aim of this section is to show that this is not a pure 
coincidence---there are general techniques which
allow to (polynomially) reduce bisimilarity to simulation
preorder/equivalence over many classes of infinite-state systems.
The material presented in this section is based mainly on
\cite{KM:Why-bisim-sim}. 

\begin{figure}[http]
\centering
{\scriptsize
\psfrag{s}[c][c]{$s$}
\psfrag{t}[c][c]{$t$}
\psfrag{la}[l][l]{$a$}
\psfrag{ra}[r][r]{$a$}
\psfrag{b}[r][r]{$b$}
\psfrag{c}[l][l]{$c$}
\psfrag{tb}[r][r]{$t_b$}
\psfrag{tc}[l][l]{$t_c$}
\psfrag{A-b}[r][r]{$\Act {\setminus} \{b\}$}
\psfrag{A-c}[l][l]{$\Act {\setminus} \{c\}$}
\psfrag{Act}[c][c]{$\Act$}
\includegraphics{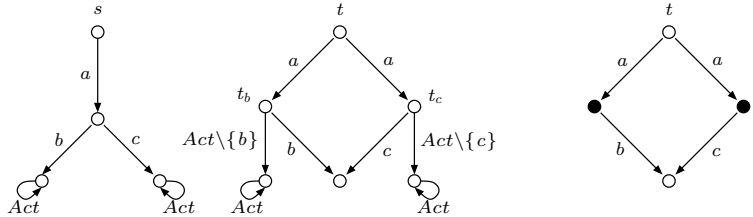}
}
\caption{The defender can enforce $b$ or $c$ in the second round.}
\label{fig-sim-force}
\end{figure}

We start with a simple observation about a specific power of the defender in
simulation games. Although the defender moves only his token during
a play, his choice of a defending move can indirectly 
``force'' the attacker to do a specific transition (with the
attacker's token) in the
next round. To illustrate this, we consider the
first two rounds of the simulation game for the states $s$ and $t$
in the transition system of Fig.~\ref{fig-sim-force} (left and
middle). 
After the attacker plays his only $a$-move, the defender can
choose between moving to $t_b$ or $t_c$. When he moves to $t_b$,
he forces
the attacker to use a $b$-move in the next round---if the
attacker plays any other action, the defender
moves to a state which enables all actions forever and therefore
wins. Similarly, when the defender moves to $t_c$, he forces the
attacker to use a $c$-move.
We say that the $b$- and $c-$ 
transitions are \emph{enforced}
by $t_b$ and $t_c$, respectively. To simplify our figures, we indicate
the states which enforce the actions of their out-going transitions
by black-filled circles. So, the middle part of Fig.~\ref{fig-sim-force}
can be simplified to the right-hand part of Fig.~\ref{fig-sim-force}.

The defender's ability to enforce the next attacker's transition 
is a crucial ingredient of several ``hardness proofs'' for simulation 
preorder/equivalence. (We address this issue in greater detail in
Section~\ref{sec-undec-lower} where we also deal with a similar technique 
for bisimilarity). 
Moreover, this was used in \cite{KM:Why-bisim-sim}
to show that there are general ``reduction schemes'' allowing for efficient
reductions of the 
$\mathbf{A} \sim \mathbf{B}$ problem to the $\mathbf{A} \sm \mathbf{B}$
problem for certain process classes $\textbf{A}$ and $\textbf{B}$. 
More specifically, such a ``reduction scheme'' defines for every 
pair of processes $s,t$ a new pair of processes $s',t'$ so that
$s \sim t$ iff $s' \sm t'$. The scheme is ``applicable'' to process
classes $\mathbf{A}$ and $\mathbf{B}$ if for all processes 
$s \in \mathbf{A}$ and $t \in \mathbf{B}$ we have that the
$s'$ and $t'$ are efficiently 
definable in the syntax of $\mathbf{A}$ 
and $\mathbf{B}$, respectively. 

\begin{figure}[http]
\centering
{\scriptsize
\psfrag{s}[c][c]{$s$}
\psfrag{t}[c][c]{$t$}
\psfrag{sb}[l][l]{$s'$}
\psfrag{tb}[c][c]{$t'$}
\psfrag{s1}[c][c]{$s_1$}
\psfrag{t1}[c][c]{$t_1$}
\psfrag{s2}[c][c]{$s_2$}
\psfrag{t2}[c][c]{$t_2$}
\psfrag{lt1}[l][l]{$t_1$}
\psfrag{rt2}[r][r]{$t_2$}
\psfrag{la}[l][l]{$a$}
\psfrag{ra}[r][r]{$a$}
\psfrag{lad}[l][l]{$a,\delta^a_1$}
\psfrag{rad}[r][r]{$a,\delta^a_2$}
\psfrag{tick}[l][l]{$\checkmark$}
\psfrag{ld1}[l][l]{$\delta_1$}
\psfrag{rd1}[r][r]{$\delta_1$}
\psfrag{ld2}[l][l]{$\delta_2$}
\psfrag{rd2}[r][r]{$\delta_2$}
\psfrag{ld1a}[l][l]{$\delta^a_1$}
\psfrag{rd1a}[r][r]{$\delta^a_1$}
\psfrag{ld2a}[l][l]{$\delta^a_2$}
\psfrag{rd2a}[r][r]{$\delta^a_2$}
\psfrag{d3}[c][c]{$\delta^a_3$}
\psfrag{lambdas}[r][r]{$\lambda^a_1,\lambda^a_2,\lambda^a_3$}
\psfrag{deltas}[l][l]{$\delta^a_1,\delta^a_2,\delta^a_3,\lambda^a_3$}
\psfrag{all}[r][r]{$\Act'$}
\psfrag{l1}[r][r]{$\lambda^a_1$}
\psfrag{l2}[l][l]{$\lambda^a_2$}
\psfrag{st}[c][c]{$s' = (s,t)$}
\psfrag{bst}[c][c]{$t' = (\bar{s},\bar{t})$}
\psfrag{rl1}[r][r]{$\lambda_1$}
\psfrag{rl2}[r][r]{$\lambda_2$}
\psfrag{ll1}[l][l]{$\lambda_1$}
\psfrag{ll2}[l][l]{$\lambda_2$}
% \psfrag{rd1}[r][r]{$\delta_1$}
% \psfrag{rd2}[r][r]{$\delta_2$}
% \psfrag{ld1}[l][l]{$\delta_1$}
% \psfrag{ld2}[l][l]{$\delta_2$}
\psfrag{s1t}[r][r]{$(s_1,t)$}
\psfrag{s2t}[r][r]{$(s_2,t)$}
\psfrag{st1}[l][l]{$(s,t_1)$}
\psfrag{st2}[l][l]{$(s,t_2)$}
\psfrag{s1t1}[c][c]{$(s_1,t_1)$}
\psfrag{s2t1}[c][c]{$(s_2,t_1)$}
\psfrag{s1t2}[c][c]{$(s_1,t_2)$}
\psfrag{s2t2}[c][c]{$(s_2,t_2)$}
\psfrag{bs1t1}[c][c]{$(\bar{s}_1,\bar{t}_1)$}
\psfrag{bs2t1}[c][c]{$(\bar{s}_2,\bar{t}_1)$}
\psfrag{bs1t2}[c][c]{$(\bar{s}_1,\bar{t}_2)$}
\psfrag{bs2t2}[c][c]{$(\bar{s}_2,\bar{t}_2)$}
\includegraphics{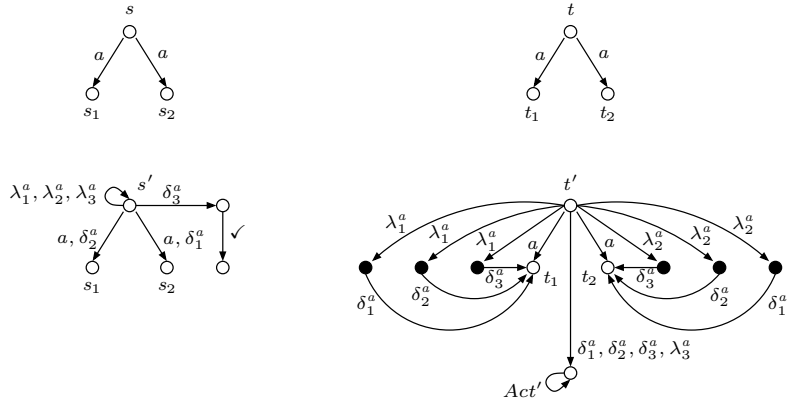}
}
\caption{The reduction of bisimilarity to simulation preorder.
  The systems $\S$ and $\T$ are in the first row (left and right, resp.), 
  and the systems $\S'$ and $\T'$ are in the second row (left and right, 
  resp.).}
\label{fig-sim-bisim}
\end{figure}

The existing reduction schemes are based on a possibility to emulate 
one round of the bisimulation game by one or two rounds of the simulation
game. Here, the above discussed enforcing of transitions is used
to emulate the ``exchange of tokens'' which can take place in the
bisimulation game. To get a better idea on how this can be done,
consider two states $s,t$ of transition systems $\S$ and $\T$ which have
the same set of actions $\Act$ and 
$\max \{d(\S),d(\T)\} \leq 3$ (i.e., the branching degrees are 
at most~3). Further,
let us suppose that $s$ and $t$ have just two successors 
$s_1,s_2$ and $t_1,t_2$, 
respectively (see top of Fig.~\ref{fig-sim-bisim}). We show 
how to emulate one round of the bisimulation game initiated in $s$ and $t$ 
by at most two rounds of the simulation game initiated in (other) states
$s'$ and $t'$ of transition systems $\S'$ and $\T'$ so that
$s \sim t$ iff $s' \sm t'$.

Here the systems $\S'$ and $\T'$ (see Fig.~\ref{fig-sim-bisim})
are obtained just by extending $\S$ and $\T$ by other
states and transitions labeled by fresh actions
(the set of actions 
of $\S'$ and $\T'$ is denoted by $\Act'$). The definition 
of $\S'$ (or $\T'$) depends just on $\S$ (or $\T$), $\Act$, and 
$\max\{d(\S),d(\T)\}$.  
The rules of the bisimulation game allow the attacker
to choose his token at the beginning of every round. If he plays
with the token put on $s$ (e.g., by performing
$s \tran{a} s_1$), the emulation is trivial
and takes just one round of the simulation game initiated in $s'$ and $t'$
(in our case, the attacker would play $s' \tran{a} s'_1$ and the defender
could also just mimic the response from the bisimulation game between 
$s$ and $t$).
Now suppose that the attacker takes the other token and plays, e.g.,
$t \tran{a} t_2$. In this case, the emulation is slightly more complicated
and takes two rounds. First, the attacker performs the 
$\lambda^a_2$-loop on $s'$. By doing so, he in fact says 
that he wants to emulate the second $a$-transition of $t$ in 
$\T$ (hence, the $\lambda$ has $a$ and $2$ as its upper and lower 
index, respectively). To enable that the attacker can emulate moves
from any state (not just $t$), we provide
 $\max\{d(\S),d(\T)\}$ distinct $\lambda^x_i$-loops 
for each action $x \in \Act$.  In Fig.~\ref{fig-sim-bisim} we indicated 
just those successors of $s'$ and $t'$ which handle the action $a$; 
if there was another $b \in \Act$, there would be 
a family of analogously constructed $\lambda^b_i$ and $\delta^b_i$ transitions 
of $s'$ and $t'$ even if $s$ and $t$ have no outgoing $b$-transitions.
As a response to the $\lambda^a_2$-loop played by the attacker, 
the defender can choose a state which enforces either $\delta^a_1$, 
$\delta^a_2$, or $\delta^a_3$. 
Intuitively, he says that he wants 
to emulate the move to the first/second/third
$a$-successor of $s$ in $\S$. The $\delta^a_3$ is needed because
the defender must be able to act accordingly for any position of the
attacker's token. This finishes the first round, i.e., the first 
emulation phase where each of the two players makes his choice. 
The purpose of the second round is to ensure that the resulting 
position of tokens (after performing the second round) really 
corresponds to the choice which has been made. In our scenario,
the attacker is forced to play the
chosen $\delta^a_i$ action; and the only possibility available to the
defender is to go to the state which was previously selected by the
$\lambda^a_2$ action, i.e., to $t'_2$. 

If one of the two players cheats in the first round by trying to
emulate a transition which does not 
really exist in $s$ or $t$, the other player wins. For example, if 
the attacker performs the $\lambda^a_3$-loop on $s'$ (i.e., he chooses
the third $a$-successor of $t$ which does not exist), 
the defender can respond by going to
a state which can simulate everything. Similarly, if the attacker plays
$\lambda^a_1$ and the defender enforces $\delta^a_3$, the 
attacker wins in two rounds by performing $\delta^a_3$ and then 
$\checkmark$. It follows that $s \sim t$ iff $s'\sm t'$. 

The above scheme is applicable to process classes $\textbf{A}$ and
$\textbf{B}$ if the syntax of $\textbf{A}$ and $\textbf{B}$ 
allows to ``test for non-enabledness'' of transitions. Examples include
PDA, BPA, OC-A, \mbox{1-safe} Petri nets, finite-state automata, etc. 
This means that, e.g., the problem $\PDA \sim \FS$ is polynomially
reducible to $\PDA \sm \FS$ and $\FS \sm \PDA$.  Moreover, simulation
preorder is easily reducible to simulation equivalence as follows: given
processes $s$ and $t$, we define other processes $s'$ and $t'$ which
have (exactly) the transitions $s' \tran{a} s$, $s' \tran{a} t$, and
$t' \tran{a} t$. We see that $s \sm t$ iff $s' \sme t'$. This
reduction is easily applicable to almost all process classes (thus, 
e.g., $\PDA \sim \FS$ is polynomially reducible to $\PDA \sme \FS$). 
However, there are also process classes to which the above scheme
is not applicable. For example, general Petri nets cannot test a place for
non-emptiness and therefore we cannot implement the families of
$\lambda$ and $\delta$ transitions in the syntax of Petri nets.
However, the bisimilarity problem for Petri nets is still polynomially
reducible to the problem of simulation preorder/equivalence by
employing a \emph{different} reduction scheme (also presented
in \cite{KM:Why-bisim-sim}). 
There are also models (like, e.g., BPP or PA) where none of the known
schemes works. An interesting question is if the existing schemes
can be further generalized so that they cover
all ``reasonable'' classes of infinite-state systems. 
A more detailed discussion can be found in \cite{KM:Why-bisim-sim}.

\subsubsection{Reducing Simulation Equivalence to Bisimilarity.}
\label{sec-sim-to-bisim}

The results which will be presented in Section~\ref{sec-results}
indicate that there cannot be any general scheme for an efficient
reduction of simulation equivalence to bisimilarity. Nevertheless,
there \emph{is} a general principle which can, in some sense, be seen
as such a ``reduction''. Of course, this ``reduction'' is not
effective in general.  It can be effectively applied only in some
restricted cases.  Nevertheless, it also reveals an interesting
relationship between simulation equivalence and bisimilarity and
therefore we present it shortly. This subsection is based
on~\cite{KM:simulation-FS-IC}.

Let $\T = (S,\Act,\tran{})$ be an image-finite transition system. A
transition $s \tran{a} t$ is \emph{maximal} iff for every transition
of the form $s \tran{a} t'$ we have that if $t \sm t'$ then also $t'
\sm t$. In other words, $s \tran{a} t$ is maximal if $t$ is maximal
w.r.t.{} simulation preorder among all $a$-successors of $s$. Note
that if the set of all $a$-successors of $s$ is nonempty, there must
be at least one maximal $a$-transition from $s$ because $\T$ is
image-finite. For example, the only maximal transition 
of the process $u$ of Fig.~\ref{fig-exa-equiv} is the middle one.
\begin{definition}
  Let $\T = (S,\Act,\tran{})$ be an image-finite transition 
  system. We define the system $\bar{\T} = (\bar{S}, \Act, \ttran{})$
  where $\bar{S} = \{\bar{s} \mid s \in S\}$ and
  $\bar{s} \ttran{a} \bar{t}$ iff $s \tran{a} t$ is a maximal transition
  of $\T$.
\end{definition}
Hence, $\bar{\T}$ is obtained from $\T$ by renaming its
states and deleting all non-maximal transitions. Now consider a
simulation game between states $s$ and $\bar{s}$. Intuitively, none of
the two players can gain anything by using the non-maximal transitions
because they are surely not the most optimal attacks/defenses. Thus,
we obtain that $s \sme \bar{s}$ for every $s \in S$. From this we
immediately get that $s \sme t$ iff $\bar{s} \sme \bar{t}$ for all
$s,t \in S$. Finally, note that if $\bar{s} \sme \bar{t}$ then also
$\bar{s} \sim \bar{t}$. To see this, one can readily check that the
relation $R = \{(\bar{s},\bar{t}) \mid \bar{s} \sme \bar{t}\}$ is a
bisimulation. As a simple consequence of presented observations, we
obtain
\begin{theorem}
\label{thm-sim-bisim}
  Let $\T$ be an image-finite transition system. For all
  $s,t \in S$ we have that $s \sme t$ iff $\bar{s} \sim \bar{t}$,
  where $\bar{s}$ and $\bar{t}$ are the ``twins'' of $s$ and $t$ in
  $\bar{\T}$, respectively.
\end{theorem}
Using the previous theorem one can ``reduce'' certain simulation
problems to their bisimulation counterparts. For example, instead of
deciding simulation equivalence between $s$ and $t$, we can (in
principle) decide bisimilarity between $\bar{s}$ and $\bar{t}$.
However, this ``reduction'' is rarely effective. If $\T$ is generated
by a PRS $\Delta$, one cannot compute another PRS $\bar{\Delta}$ which
generates the system $\bar{\T}$ in general. It is not even clear if
such a $\bar{\Delta}$ exists. Nevertheless, the effective construction
is possible in some restricted cases. For example, if $\Delta$ is
deterministic, then trivially $\bar{\Delta} = \Delta$.  If $\Delta$ is
a FS system, then $\bar{\Delta}$ is constructible in polynomial time
because simulation preorder between the states of $\T_\Delta$ is
computable in polynomial time. A less trivial example are OC-N
systems---if $\Delta$ is an OC-N system, then $\bar{\Delta}$ is an
effectively definable OC-A system \cite{JKM:one-counter-sim-bisim}.
Hence, certain simulation problems for OC-N processes are effectively
reducible to the corresponding bisimulation problems over OC-A
processes, and the decidability of some of them has indeed been
established in this way~\cite{JKM:one-counter-sim-bisim}.

\subsection{Decidability and Upper Complexity Bounds}
\label{sec-dec-upper}

\subsubsection{Bisimulation Bases.}
\label{sec-bases}

The technique of \emph{bisimulation bases} was pioneered by 
Caucal in \cite{Caucal:selfbisimulation-RAIRO}. We start by 
explaining the underlying principle which is to some extent 
model-independent. The introduced notions are then illustrated
on a concrete example. Finally, we show how the method applies
to weak bisimilarity. 

Since the ``classical'' results about bisimulation bases are carefully 
presented in \cite{BCMS:Infinite-Structures-HPA}, we mention them 
just shortly. The main point
of this section is the part about weak bisimilarity which is based on 
recent results \cite{KM:BPA-nBPP-FS-weak-TCS}. 

\begin{definition}
\label{def-strong-expand}
Let $\T_1 = (S_1,\Act,\tran{}_1)$ and $\T_2 = (S_2,\Act,\tran{}_2)$ be
two transition systems; we will write just $\tran{}$ instead of
$\tran{}_1$, $\tran{}_2$.  Let $R \subseteq S_1 {\times} S_2$.  
We say that a \emph{pair} $(s,t) \in S_1 \times S_2$ \emph{expands in} $R$ if
  \begin{itemize}
  \item for every  $s \tran{a} s'$ there is some 
     $t \tran{a} t'$ such that $(s',t') \in R$;
  \item for every  $t \tran{a} t'$ there is some 
     $s \tran{a} s'$ such that $(s',t') \in R$.
  \end{itemize}
Now let $P,R \subseteq S_1 {\times} S_2$. We say that 
$P$ \emph{expands in} $R$ if all pairs of $P$ expand in $R$.
\end{definition}
Let $\mathbf{C_1}$ and $\mathbf{C_2}$ be subclasses of process 
rewrite systems (not necessarily different), and let 
$\Delta_1 \in \mathbf{C_1}$ and $\Delta_2 \in \mathbf{C_2}$.
Further, let 
\[
  \Bis = \{(\alpha,\beta) \mid \alpha \in \T_{\Delta_1}, \beta \in 
           \T_{\Delta_2}, \alpha \sim \beta \}
\]
be the bisimilarity relation between the processes of $\Delta_1$ and
$\Delta_2$. A \emph{bisimulation base $\B$} (for $\Delta_1$ and $\Delta_2$)
is a finite subset of
$\Bis$ consisting only of ``crucial'' bisimilar pairs from which the
whole relation $\Bis$ can be generated in some ``syntactic'' way.
More precisely, one defines an operator $\Gen$ which for each relation 
$R \subseteq \T_{\Delta_1} \times \T_{\Delta_2}$ returns another relation
$\Gen(R) \subseteq \T_{\Delta_1} \times \T_{\Delta_2}$ so that the
following conditions are satisfied: 
\begin{itemize}
\item[(1)] $\Gen(\B) = \Bis$.
\item[(2)] $\Gen$ is monotonic, i.e., if $R \subseteq R'$ then 
  $\Gen(R) \subseteq \Gen(R')$.
\item[(3)] If $R$ is a relation which expands in $\Gen(R)$, then also 
   $\Gen(R)$ expands in $\Gen(R)$. (In other words, if $R$ expands in 
   $\Gen(R)$ then $\Gen(R)$ is a bisimulation.)  
\end{itemize}

Of course, finite bisimulation bases, and the associated $\Gen$
operators, exist only for \emph{some} subclasses $\mathbf{C_1}$ and
$\mathbf{C_2}$ of PRS. If the question whether $(\alpha,\beta)
\in\Gen(R)$ is semidecidable ($R$ being finite), then the question
whether $R$ expands in $\Gen(R)$ is also semidecidable. Therefore, the
problem $\mathbf{C_1} \sim \mathbf{C_2}$ is semidecidable---to verify
that $\alpha\sim\beta$, we can run a semidecision procedure which is
guaranteed to find a finite relation $R$ which expands in $\Gen(R)$
and for which $(\alpha,\beta) \in \Gen(R)$ (on condition that such
a relation $R$ exists).  
If $\alpha \sim \beta$, then this procedure halts because
the finite base $\B$ must eventually be found (observe that $\B$ has
all the required properties).  And if the procedure halts because
some relation $R$ satisfying all of the required properties is found, 
we can conclude that $\Gen(R)$ is a bisimulation (due to~(3) above), 
hence $\alpha \sim \beta$.

Since the negative subcase $\mathbf{C_1} \not\sim \mathbf{C_2}$ is
semidecidable due to generic reasons (see
Theorem~\ref{thm-finitely-branching}), we in fact obtain the
decidability of the $\mathbf{C_1} \sim \mathbf{C_2}$ problem.
 
Now assume that the membership in $\Gen(R)$ is even \emph{decidable}
for every $R$, and that for all $\Delta_1$ and $\Delta_2$ there is an
effectively computable relation $\G$ which is guaranteed to subsume
the base.  Then the base is computable by the algorithm of
Fig.~\ref{fig-alg-B}.  Note that if $\B \subseteq R$, then $\B$
expands in $\Gen(R)$, because $\B$ expands in $\Gen(\B)$ and $\Gen$ is
monotonic (see~(2) above). This means that $\B \subseteq B$ is an
invariant of the \textbf{repeat-until} loop of the algorithm of
Fig.~\ref{fig-alg-B}.  Moreover, if $\G$ is computable in polynomial
time (in the size of $\Delta_1$ and $\Delta_2$), and the membership in
$\Gen(R)$ is decidable in polynomial time, then the base is computable
in polynomial time.

\begin{figure}[http]
\begin{tabbing}
\hspace*{1em} \=  \hspace*{1em} \= \hspace*{1em} \= \kill
\makebox[4.5em][l]{\textbf{Input:}} Process Rewrite Systems 
   $\Delta_1 \in \mathbf{C_1}$, $\Delta_2 \in \mathbf{C_2}$.\\
\makebox[4.5em][l]{\textbf{Output:}} The base $\B$.\\[1ex]
\> $B := \G;$\\
\> \kw{repeat}\\
\>\> $R := B;$ $B := \emptyset$\\
\>\> \kw{for all } $(\alpha,\beta) \in R$ \kw{ do}\\
\>\>\> \kw{if} $(\alpha,\beta)$ expands in $\Gen(R)$ \kw{then}
 $B := B \cup \{(\alpha,\beta)\}$ \kw{fi}\\
\>\> \kw{od};\\
\> \kw{until } $B = R$\\
\> $\B := B$;
\end{tabbing}
\caption{An algorithm for computing $\B$}
\label{fig-alg-B}
\end{figure}

\begin{example}
If $\mathbf{C_1} = \mathbf{C_2} = \nBPA$ and 
$\Delta_1 = \Delta_2 = \Delta$, one can put
\[
  \B = \{(X,\alpha) \mid X \in \C(\Delta), \alpha \in \S(\Delta), 
         X \sim \alpha\}
\]
and $\Gen(R) = \Congr(R)$, where $\Congr(R)$ is the least congruence
over $\S(\Delta)$ w.r.t.\ ``$\cdot$'' subsuming $R$. The $\B$ can be
over-approximated by a finite relation
\[
  \G = \{(X,\alpha) \mid X \in \C(\Delta), \alpha \in \S(\Delta), 
          \norm(X) = \norm(\alpha)\}
\]
where $\norm(\alpha)$ is the length of the shortest sequence $w \in
\Act^*$ such that $\alpha \tran{w} \varepsilon$.  Realize that $\B$
and $\G$ are finite relations because bisimilar processes must have
the same norm and there are only finitely many processes with a given
finite norm.

To get some idea on how all this works, let us prove that $\Gen(\B) =
\Bis$.  Clearly $\Gen(\B) \subseteq \Bis$, because bisimilarity is a
congruence over $\S(\Delta)$ w.r.t.\ ``$.$''. To prove $\Bis \subseteq
\Gen(\B)$, consider some $\alpha \sim \beta$; by induction on
$\norm(\alpha) = \norm(\beta)$ we prove that $(\alpha,\beta) \in
\Gen(\B)$.  If $\norm(\alpha) =1$, then $\alpha = X$ for some $X$ and
hence $(\alpha,\beta) \in \B$.  Now let $\norm(\alpha) > 1$. Then
$\alpha = X . \gamma$ and $\beta = Y . \delta$; let us assume that
$\norm(X) \leq \norm(Y)$ (the other case is symmetric).  Let $X .
\gamma \tran{w} \gamma$ where $\length(w) = \norm(X)$.  The bisimilar
process $Y . \delta$ must be able to match this sequence of
transitions by some $Y . \delta \tran{w} \xi . \delta$ so that $\gamma
\sim \xi . \delta$. Observe that $(\gamma, \xi . \delta) \in \Gen(\B)$
by induction hypothesis.  As $X . \gamma \sim Y . \delta$ and $\gamma
\sim \xi . \delta$, we also have $X . \xi . \delta \sim Y . \delta$
and thus $X . \xi \sim Y$ by applying the right cancellation law which
is admitted by normed BPA processes. This means that $(Y,X . \xi) \in
\B$. To sum up, $(\gamma, \xi . \delta) \in \Gen(\B)$ and $(Y,X . \xi)
\in \B$, which means that also $(X . \gamma, Y . \delta) \in
\Gen(\B)$.

The operator $\Gen$ is clearly monotonic, and one can show that
the condition~(3) above is also satisfied.
\end{example}

From the previous example, it follows that the problem $\nBPA \sim
\nBPA$ is decidable. This proof is essentially due to Caucal
\cite{Caucal:selfbisimulation-RAIRO}.  Later, the structure of $\B$
was further simplified so that its size (and the size of $\G$) became
\emph{polynomial} in the size of $\Delta$, and a suitable $\Gen$ was
designed so that the algorithm of Fig.~\ref{fig-alg-B} terminates in
\emph{polynomial time} \cite{HJM:BPA-polynomial-TCS}. Hence, $\nBPA
\sim \nBPA$ is in $\PTIME$.  In \cite{CHHS:BPA-bisimilarity-IC}, it
has been shown that a finite bisimulation base exists also for general
(not necessarily normed) BPA processes. This implies the
semidecidability (and hence also the decidability) of the $\BPA \sim
\BPA$ problem. An algorithm for computing the bisimulation base for
general BPA processes appeared in \cite{BCS:BPA-bisim-elementary}, and
this result led to an elementary upper complexity bound for the $\BPA
\sim \BPA$ problem (a later result due to Srba \cite{Srba:BPA-PSPACE}
shows that the problem is \PSPACE-hard).

Finite bisimulation bases exist also for BPP processes
\cite{CHHM:BPP-bisimilarity}. In the case of normed BPP processes, the
base is small and can be computed in polynomial time
\cite{HJM:nBPP-polynomial-MSCS}.  The general problem $\BPP \sim \BPP$
is \PSPACE-hard \cite{Srba:BPP-PSPACE}, and in fact \PSPACE-complete
\cite{Jancar:BPP-PSPACE} (see also Section~\ref{sec-DD-functions}).

The technique of bisimulation bases works also for weak bisimilarity,
if the notion of expansion is modified as follows:

\begin{definition}
\label{def-weak-expand}
  Let $\T_1 = (S_1,\Act,\tran{})$ and $\T_2 = (S_2,\Act,\tran{})$
  be transition systems, and let
  $R \subseteq S_1 {\times} S_2$ be relations. A pair 
  $(s,t) \in S_1 \times S_2$ \emph{weakly expands} in $R$ if
  \begin{itemize}
  \item for every  $s \tran{a} s'$ there is some 
     $t \Tran{a} t'$ such that $(s',t') \in R$;
  \item for every  $t \tran{a} t'$ there is some 
     $s \Tran{a} s'$ such that $(s',t') \in R$.
  \end{itemize}
  Let $P,R \subseteq S_1 {\times} S_2$. We say that 
  $P$ \emph{weakly expands in} $R$ if all pairs of $P$ weakly expand in $R$.
\end{definition}
The ``asymmetry'' which appears in the definition of weak expansion
matches the original definition of weak bisimilarity used in
\cite{Milner:book}. The principle would work also for the ``symmetric
version'' of weak expansion, but the introduced asymmetry leads to
important algorithmic simplifications.

\begin{example}
\label{exa-BPA-FS}
Let $\mathbf{C_1} = \BPA$, $\mathbf{C_2} = \FS$, 
$\Delta$ be a BPA system and $\Delta_2$ a FS system such that
$\C(\Delta) \cap \C(\Delta_2) = \emptyset$. For technical convenience,
we put $\Delta_1 = \Delta \cup \Delta_2$. Note that $\Delta_1$
is a BPA system. Now let 
\begin{eqnarray*}
  \B & = & \{(AX,Y) ~\mid~ A \in \C(\Delta),\  X,Y \in \C(\Delta_2),\ 
      AX \wsim Y \}\\
      & \cup & \{(A,Y) ~\mid~ A \in \C(\Delta),\ Y \in \C(\Delta_2),\
      A \wsim Y \}\\
      & \cup & \{(\varepsilon,Y) ~\mid~ Y \in \C(\Delta_2),\ 
        \varepsilon \wsim Y \}
\end{eqnarray*}
Note that $\B$ can be over-approximated by a relation $\G$ of size
$\calO(|\Delta_1| \cdot |\Delta_2|^2)$ which consists of all
syntactically conformable pairs.

For every relation $R \subseteq \G$ we define $\Gen(R)$ to be
the least relation $K$ (between states of $\T_{\Delta_1}$
and states of $\T_{\Delta_2}$) subsuming $R$ such that 
\begin{itemize}
\item whenever $(\alpha X,Y) \in K$ and $(\beta, X) \in K$, then also 
  $(\alpha\beta, Y) \in K$;
\item whenever $(\beta, X) \in K$ where $\norm(\beta) = \infty$, then
  also $(\beta\gamma, X) \in K$ for all $\gamma \in \S(\Delta_1)$.
\end{itemize}
One can readily check that $\Gen(\B) = \Bis$
and that $\Gen$ is monotonic. The proof that the condition~(3) is also
satisfied is more involved and can be found in \cite{KM:BPA-nBPP-FS-weak-TCS}.

Since the membership in $\Gen(R)$ is easily decidable in polynomial
time, one is tempted to conclude that the algorithm of
Fig.~\ref{fig-alg-B} computes the base in polynomial time. This is
indeed the case, but an additional problem has to be solved first.
Let us consider, e.g., a pair of the form $(A,Y)$ where $A \in
\C(\Delta)$ and $Y \in \C(\Delta_2)$.  According to
Definition~\ref{def-weak-expand}, $(A,Y)$ weakly expands in $\Gen(R)$
if for every ``$\tran{a}$'' move of one of the two processes there is a
``$\Tran{a}$'' move of the other process such that the resulting pair
belongs to $\Gen(R)$. The problem is that $A$ can have
\emph{infinitely many} $\Tran{a}$ successors and hence we cannot
simply try them one by one. If we denote $\Reach^a_A = \{\alpha \mid A
\Tran{a} \alpha\}$ and $\Gen_X(R) = \{\alpha \mid (\alpha,X) \in
\Gen(R)\}$, the question whether for a given $Y \tran{a} X$ there is
some $A \Tran{a} \alpha$ such that $(\alpha,X) \in \Gen(R)$ reduces to
the problem of checking whether $\Reach^a_A \cap \Gen_X(R) = \emptyset$.
Since both
sets can be infinite, the key is to find a suitable finite
representation for them. In this case, it suffices to employ
finite-state automata---both sets are regular and the associated
finite-state automata are small and efficiently computable.  Now the
emptiness of $\Reach^a_A \cap \Gen_X(R)$ can be decided in polynomial
time by standard methods of automata theory \cite{HU:book}.
\end{example}

The details can be found in \cite{KM:BPA-nBPP-FS-weak-TCS}, where 
a similar method is used to show that also the problem
$\nBPP \sim \FS$ is decidable in polynomial time. In this case,
the set of states which are reachable from a given BPP process in one 
``$\Tran{a}$'' move is represented by a context-free grammar.
Since the structure of the base is still regular,
one can rely on the standard result saying that the emptiness of
the intersection of a given CF-language and a given regular language
can be decided in polynomial time. Recently, the method for
BPA and FS processes described in Example~\ref{exa-BPA-FS} was
generalized to PDA and FS systems and other behavioral equivalences
\cite{KM:generic-PDA-FS}. In \cite{BKS:pBPA-pBPP-bisim}, it is shown 
that the technique of bisimulation bases is applicable also to 
probabilistic bisimilarity and probabilistic extensions of 
BPA, BPP, and PDA processes.

\subsubsection{Characteristic Formulae for Finite-State Processes.}
\label{sec-equiv-FS}

The problem of checking a given behavioral equivalence between an
infinite-state process $g$ and a finite-state specification $f$ has
recently been identified as an important subcase of the general
equivalence-checking problem.  There are two main reasons why this
question attracts a special attention. First, in equivalence-based
verification, one usually compares a ``real-life'' system with an
abstract behavioral specification.  A faithful model of the real-life
system often requires features like counters, or subprocess creation,
or unbounded buffers, that make the model infinite-state. On the other
hand, the behavioral specification is usually abstract, hence
naturally finite-state. Moreover, infinite-state systems are often
abstracted to finite-state systems even before applying further
analytical methods.  This approach naturally subsumes the question if
the constructed abstraction is correct (i.e., equivalent to the
original system).  The second reason is that checking equivalence
between an infinite and a finite-state process is computationally
easier than comparing two infinite-state processes (as also
demonstrated by results of Section~\ref{sec-results}).

In this section we first recall the notion of a characteristic formula
and show how to construct characteristic formulae in the modal
$\mu$-calculus \cite{SI:char-formulae}. Then, we concentrate on
bisimulation-like equivalences. We present a simple theorem which
reformulates the problem of bisimilarity between an infinite and a
finite-state process to some kind of ``reachability question''.  This
approach originated in
\cite{JM:PN-properties,AK:LCSvsFS,JK:FS-bisim-ENTCS}. A more abstract
formulation which applies also to weak bisimilarity is due to
\cite{JKM:bisim-like-TCS}. Using this result, we show that
characteristic formulae for finite-state systems w.r.t.\ 
bisimulation-like equivalences can also be constructed in the
branching-time logic EF. This logic is much simpler than the modal
$\mu$-calculus, and consequently the model-checking problem with the
logic EF is decidable for many classes of infinite-state systems.
Thus, a number of decidability/complexity results about checking
bisimilarity between infinite and finite-state processes have been
obtained \cite{JKM:bisim-like-TCS}.
 
\begin{definition}
  Let $\F = (F,\Act,\tran{})$ be a finite-state system, $f \in F$, and
  $\leftrightarrow$ an equivalence over the class of all processes.
  Let $\mathbf{C}_f$ be the class of all processes $s$ such that the
  set of actions of $s$ (in its underlying transition system) is
  included in $\Act$.  A formula $\varphi$ is \emph{characteristic}
  for $f$ w.r.t.{} $\leftrightarrow$ if for every $s \in \mathbf{C}_f$
  we have that $s \leftrightarrow f$ iff $s$ satisfies $\varphi$.
\end{definition}
Characteristic formulae w.r.t.\ $\sim_i$ (for given $i \in \Nseto$ and
$\Act$) are easily definable in \emph{Hennessy-Milner (H.M.) logic}
\cite{Milner:book}.  The syntax of H.M.\ logic is given by
\[
  \varphi \quad ::= \quad \ttt \quad  
                    | \quad \varphi \wedge \varphi \quad 
                    | \quad \neg \varphi \quad 
                    | \quad \Ex{a} \varphi 
\]
where $a$ ranges over actions.  Formulae are interpreted over
processes; the propositional connectives have the standard meaning and
$s \models \Ex{a} \varphi$ iff there is some $s \tran{a} t$ such that
$t \models \varphi$. A formula $\neg \Ex{a} \neg \varphi$ is usually
abbreviated to $\Al{a} \varphi$.

\begin{figure}
\centering
\psfrag{a}[r][r]{$a$}
\psfrag{ab}[c][c]{$a,b$}
\psfrag{b}[c][c]{$b$}
\psfrag{s}[c][c]{$f$}
\psfrag{t}[c][c]{$h$}
\includegraphics{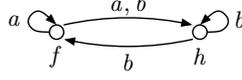}
\caption{Processes $f$ and $h$.}
\label{fig-finite-f}
\end{figure}

Now consider the transition system of Fig.~\ref{fig-finite-f}.
The behavior of $f$ and $h$ is described (up to bisimilarity) by the 
following recursively defined properties $\varphi_f$ and $\varphi_h$, 
respectively.
\begin{eqnarray*}
\varphi_f & \ \equiv \ & \Ex{a} \varphi_f\ \wedge\  \Ex{a} \varphi_h\ \wedge\  
                     \Ex{b} \varphi_h\ \wedge\ \Al{a}(\varphi_f \vee \varphi_h)
                     \ \wedge\ \Al{b} \varphi_h\\
\varphi_h & \ \equiv \ & \Ex{b} \varphi_f\ \wedge\ \Ex{b} \varphi_h\ \wedge\ 
                     \Al{a}\fff\ \wedge\ \Al{b}(\varphi_f \vee \varphi_h)
\end{eqnarray*}
These equations can be used to construct characteristic formulae for $f$
and $h$ w.r.t.{} $\sim_i$; we inductively define the
family of $\xi_i^f$ and $\xi_i^h$ formulae as follows: 
\[
\begin{array}{lclclcl}
  \xi_0^f & = & \ttt & \hspace*{3em} & \xi_0^h & = & \ttt\\
  \xi_{i+1}^f & = & \varphi_f[\xi_i^f/\varphi_f,\xi_i^h/\varphi_h] &&
  \xi_{i+1}^h & = & \varphi_h[\xi_i^f/\varphi_f,\xi_i^h/\varphi_h]
\end{array}
\]
Here $\varphi[\xi/\psi]$ denotes the formula obtained from $\varphi$
by replacing each occurrence of subformula $\psi$ with formula $\xi$.
A straightforward proof confirms that for every process $s \in
\mathbf{C}_f$ and $i \in \Nseto$ we have that $s \sim_i f$ iff 
$s \models \xi_i^f$, and $s \sim_i h$ iff $s \models \xi_i^h$.  By
Theorem~\ref{thm-finitely-branching}, this means that
$\bigwedge_{i=0}^\infty \xi_i^f$ and $\bigwedge_{i=0}^\infty \xi_i^h$
are characteristic formulae for $f$ and $h$ w.r.t.\ $\sim$,
respectively.  These infinite conjunctions can be encoded in the modal
$\mu$-calculus \cite{Kozen:mu-calculus} by translating the recursive
dependence between $\varphi_f$ and $\varphi_h$ into an explicit
greatest fixed-point definition; thus, we obtain the formula $\Phi^f$.
\begin{eqnarray*}
\Phi^f\ \equiv \ & \ \nu S. \ & \Ex{a} S\ \wedge\  \Ex{a} \varphi_h\ \wedge\  
                     \Ex{b} \varphi_h\ \wedge\ \Al{a}(S \vee \varphi_h)
                     \ \wedge\ \Al{b} \varphi_h \ 
                       \mbox{ where} \label{formula-mu}\\
        \varphi_h \ \equiv \ & \ \nu T. \ & \Ex{b} S\ \wedge\ 
                     \Ex{b} T\ \wedge\ \Al{a}\fff\ \wedge\ \Al{b} (S \vee T)
\end{eqnarray*}
An analogous construction works also for weak bisimilarity.  Instead
of the ``$\Ex{a}$'' modality of H.M.\ logic we employ its ``weak
form'' $\wEx{a}$ defined by $\wEx{a} \varphi \equiv \Diamond_\tau
\Ex{a} \Diamond_\tau \varphi$ where $s \models \Diamond_\tau \varphi$
iff there is $s \Tran{\tau} t$ such that $t \models \varphi$. Since
the ``$\Diamond_\tau$'' is expressible in the modal $\mu$-calculus,
one can construct characteristic formulae w.r.t.\ $\wsim$ in this
logic.

Characteristic formulae w.r.t.\ simulation equivalence are also easily
definable in the modal $\mu$-calculus. To see this, examine 
the recursively defined properties $\psi_f,\psi_h$ and $\varrho_f,\varrho_h$:
\[
\begin{array}{lclclcl}
\psi_f & \ \equiv \ & \Ex{a} \psi_f\ \wedge\  \Ex{a} \psi_h\ \wedge\  
                     \Ex{b} \psi_h & \hspace*{1em} &
\varrho_f & \ \equiv \ & \Al{a}(\varrho_f \vee \varrho_h)
                     \ \wedge\ \Al{b} \varrho_h\\
\psi_h & \ \equiv \ & \Ex{b} \psi_f\ \wedge\ \Ex{b} \psi_h & &
\varrho_h & \ \equiv \ & \Al{a}\fff\ \wedge\ 
                     \Al{b} (\varrho_f \vee \varrho_h)
\end{array}
\]
A closer look reveals that for every $s \in \textbf{C}_f$ we have
$s \models \psi_f$ iff $f \sm s$, and $s \models \varrho_f$ iff
$s \sm f$. Hence, $s \sme f$ iff $s \models \psi_f \wedge \varrho_f$.
The formulae $\psi_f$ and $\varrho_f$ can be encoded in the modal
$\mu$-calculus similarly as the formula $\varphi_f$ above. 

To sum up, the modal $\mu$-calculus is sufficiently powerful to express
characteristic formulae w.r.t.\ bisimilarity and 
simulation equivalence, and the size of these formulae 
is essentially the same as the size of the underlying 
transition system of $f$. Thus, the problem of checking bisimilarity
and simulation equivalence with a finite-state process is polynomially
reducible to the model-checking problem with the modal $\mu$-calculus.
This is applicable to PDA and BPA processes where model-checking 
the modal $\mu$-calculus is known to be \EXPTIME-complete 
\cite{Walukiewicz:CAV96-IC}; hence, the problems
$\PDA \sim \FS$, $\PDA \wsim \FS$, $\PDA \sm \FS$, $\FS \sm \PDA$, 
and $\PDA \sme \FS$ are in $\EXPTIME$. The bounds for simulation 
are already tight, because these problems are also $\EXPTIME$-hard
\cite{KM:PDA-BPA-sim-bisim}. Actually, this holds even for BPA.
However, we can do better for bisimilarity; the problems 
$\PDA \sim \FS$ and $\PDA \wsim \FS$ are $\PSPACE$-complete
\cite{Mayr:PDA-FS-bis-hard,KM:PDA-BPA-sim-bisim}. 
This requires an application of a different method which is described below. 

If $\mathbf{C}$ is a class of processes such that ${\sim_{i-1}} =
{\sim_i}$ over $\mathbf{C} \times \mathbf{C}$, then $\sim_i$ is a bisimulation
relation and hence ${\sim_{i-1}} = {\sim_i} = {\sim}$ over $\mathbf{C} \times
\mathbf{C}$. For example, if $\mathbf{C}$ is the set of processes 
of a finite-state transition system with $k$ states, then surely 
${\sim_{k-1}} = {\sim_k}$ because any equivalence over $\mathbf{C}$ 
has at most $k$ equivalence classes and 
${\sim_{i+1}} \subseteq {\sim_i}$ for every $i \in \Nseto$.  
The same holds for $\wsim_i$.  The following
theorem \cite{JKM:bisim-like-TCS} presents a simple (but
important) observation about the problem of bisimilarity-checking with
finite-state processes.

\begin{theorem}
\label{thm-bis-FS}
Let $\G = (G,\Act,\tran{})$ be a (general) transition system and
\mbox{$\F = (F,\Act,\tran{})$} a finite-state transition system with $k$
states. States $g \in G$ and $f \in F$ are bisimilar iff the following
conditions hold:
\begin{itemize}
\item $g \sim_k f$;
\item for each state $g'$ such that $g \tran{}^* g'$ there is a
   state $f' \in F$ such that $g' \sim_k f'$.
\end{itemize}
\end{theorem}
\begin{proof}
``$\Longrightarrow$'' is obvious. To prove the ``$\Longleftarrow$''
direction, we show that the relation $R \subseteq G \times F$ given by
\[
  R = \{(g',f') \mid  g \rightarrow^* g' \mbox{ and } g' \sim_k f'\}
\]
is a bisimulation. Let $(g',f') \in R$ and let
$g' \tran{a} g''$ for some $a \in \Act$ (the case when
$f' \tran{a} f''$ is handled in the same
way). By definition of $\sim_k$, there is an $f''$ such that
$f' \tran{a} f''$ and $g'' \sim_{k-1} f''$.
It suffices to show that
$g'' \sim_k f''$; as $g \rightarrow^* g''$, there is a state
$\bar{f}$ of $\F$ such that $g'' \sim_k \bar{f}$. By
transitivity of $\sim_{k-1}$
we have $\bar{f} \sim_{k-1} f''$, hence
$\bar{f} \sim_k f''$
(remember that ${\sim_{k-1}} = {\sim_k}$ over $F \times F$). Now
$g'' \sim_k \bar{f} \sim_k f''$
and thus $g'' \sim_k f''$ as required. Clearly $(g,f) \in R$ and
the proof is finished.
\end{proof}
The previous theorem holds also for weak bisimilarity (we use
$\wsim_k$ instead of $\sim_k$, and $\Tran{a}$ instead of $\tran{a}$).

Theorem~\ref{thm-bis-FS} is applicable to a variety of models.
Since $\sim_k$ is decidable 
for all ``reasonably defined'' classes of processes, the problem of
bisimilarity-checking between infinite-state processes of a class $\mathbf{C}$
and finite-state processes reduces to a kind of reachability problem for
$\mathbf{C}$---all we need is an algorithm which, for a given 
process $s$ of $\mathbf{C}$, decides if $s$ can reach a state $s'$ 
which is not related by $\sim_k$ to
any state of the considered finite-state system. In some cases, this
is quite easy.

\begin{example}
  Let $p\alpha$ be a PDA process. The behavior of PDA processes up to
  $\sim_k$ is determined by the current control state and the top $k$
  symbols of the stack. Hence, for all processes $q \beta$ where the
  length of $\beta$ is bounded by $k$ we do the following (re-using
  the computational space for each of the exponentially many $q
  \beta$'s): first we decide if there is some state $f$ of the given
  finite-state system such that $q \beta \sim_k f$ (note that this can
  be done in polynomial space). If not, we either decide if $p \alpha
  \tran{}^* q \beta$ (when $|\beta| < k$), or if $p \alpha \tran{}^* q
  \beta \gamma$ for some $\gamma$ (when $|\beta| = k$). This can be done
  in polynomial time by employing standard techniques for pushdown
  automata \cite{HU:book}.  Thus, we obtain a polynomial-space
  algorithm for the problem $\PDA \sim \FS$ (the $\PSPACE$-hardness is
  due to \cite{Mayr:PDA-FS-bis-hard}).
\end{example}
Similarly, one can handle other models like
BPP, PA, or Petri nets; proofs are still simple but not completely
immediate \cite{JM:PN-properties,JK:FS-bisim-ENTCS}.

With help of Theorem~\ref{thm-bis-FS} one can also construct
characteristic formulae w.r.t.\ strong and weak bisimilarity in the
logic EF. This logic is obtained  by extending the
H.M.\ logic with the ``$\Diamond$'' (reachability) operator; 
$s \models \Diamond \varphi$ iff there is $s \tran{}^* s'$ such that
$s' \models \varphi$. For the construction of characteristic formulae
w.r.t.\ $\wsim$, we also need the aforementioned ``$\Diamond_\tau$'' 
operator to express the ``$\wEx{a}$'' 
modality. The dual operators are 
$\Box \varphi \equiv \neg \Diamond \neg \varphi$
and $\Box_\tau \varphi \equiv \neg \Diamond_\tau \neg \varphi$.
A characteristic formula $\Phi^f$ for the process $f$ of
Fig.~\ref{fig-finite-f}  
w.r.t.\ $\sim$ (or $\wsim$) in the logic EF looks as follows:
\begin{eqnarray}
  \Phi^f\ &\ \equiv\ &\ \xi_k^f \ \wedge\ \Box(\xi_k^f \vee \xi_k^h)
  \label{formula-EF}
\end{eqnarray}
Here $\xi_k^f$ and $\xi_k^h$ are characteristic formulae for $f$ and
$h$ w.r.t.{} $\sim_k$ (or $\wsim_k$). Note that, in general, the size
of the formula~(\ref{formula-EF}) is \emph{exponential} in the size of the
underlying transition system of $f$. However, the size of the
DAG\footnote{A DAG (directed acyclic graph or ``circuit'') 
representing a formula $\varphi$
is obtained from the syntax tree of $\varphi$ by identifying the nodes
corresponding to the same subformula.} representing this formula 
is only \emph{polynomial}. This is
important because the complexity of many model-checking algorithms 
depends on the size of the DAG rather then on the size of the formula
itself. Moreover, the DAG representing $\Phi^f$ is computable in polynomial
time. Thus, results about model-checking with the logic EF carry over to 
the problem of strong/weak bisimilarity with a finite-state process.
For example, model-checking the logic EF is decidable for PA processes 
\cite{Mayr-EF-TCS} (while model-checking the modal $\mu$-calculus is 
undecidable already for BPP), and thus we obtain the decidability of
$\PA \sim \FS$ and even $\PA \wsim \FS$. Since model-checking
the logic EF for PDA is \PSPACE-complete \cite{Walukiewicz:CTL-PDA}, 
we obtain that the $\PDA \sim \FS$ and $\PDA \wsim \FS$ 
problems are in \PSPACE\ and hence 
\PSPACE-complete \cite{KM:PDA-BPA-sim-bisim}.

Recently, Theorem~\ref{thm-bis-FS} and the corresponding results
about characteristic formulae have been generalized also to other 
behavioural equivalences \cite{KS:strong-reg-equiv-EF}.

\subsubsection{DD-functions}
\label{sec-DD-functions}

The technique of DD-functions was introduced
in \cite{Jancar:BPP-PSPACE} in order to show that the problem
$\BPP \sim \BPP$ is in \PSPACE. Combined with Srba's result
\cite{Srba:BPP-PSPACE}, \PSPACE-completeness has thus been established. 
The technique of DD-functions was then also used in
demonstrating the decidability of $\BPA \sim \BPP$ \cite{JKM:BPA-BPP-bisim}. 

Let $\T = (S,\Act,\tran{})$ be a transition system. 
Stipulating that $\min \emptyset = \omega$, for all $s,t \in S$ 
we define the \emph{distance from $s$ to $t$} by
\[ 
  \dist(s,t) = \min\big\{\,\length(w) \mid s \tran{w} t
\,\big\}\mbox{.}
\]
Here $\omega$ denotes an infinite amount. The set $\Nseto \cup \{\omega\}$
is denoted $\Natomega$, and we put $\omega - n = \omega$ for each 
$n \in \Natomega$. 

DD-functions are defined inductively. First, for every action $a$
we define a function $dd_a$ which, for every process $s$, gives 
the \emph{``distance to disabling''} the action $a$. Formally,
\[
  dd_a(s) = \min\big\{\,\dist(s,t) \mid t \mbox{ has no $a$-successor}\,
  \big\}\mbox{.}
\]
Given a tuple of (so far defined) DD-functions
$\F = (d_1,\ldots,d_k)$, we observe that each transition
$s \tran{a} t$ determines a \emph{change} of $\F$,
denoted $\F(t)-\F(s)$, which is a $k$-tuple of values
from $\{-1\} \cup \Natomega$ given by
\[
  \F(t)-\F(s) = \big(d_1(t) - d_1(s), \ldots, d_k(t) - d_k(s)\big)\mbox{.}
\]
Note that $d_i(s)=\omega$ implies $d_i(t)=\omega$. 
For technical reasons, we can then view $d_i(t) - d_i(s)$ as
undefined, being interested only in changes of (so far) finite
DD-functions.

The notion of change is used in the inductive step of the definition
of DD-functions. For each triple
$(a,\F,\delta)$, where $a$ is an action, $\F$ is a $k$-tuple
of DD-functions, and $\delta$ is a $k$-tuple 
of values from $\{-1\}\cup\Natomega$, the function
$dd_{(a,\F,\delta)}$ (distance to disabling the action $a$ causing
the change $\delta$ of $\F$) is also a DD-function, defined by
\[
  dd_{(a,\F,\delta)}(s) =
    \min\big\{\,\dist(s,t) \,\mid\, \forall r: \mbox{ if } t\tran{a} r
    \mbox{ then } \F(r)-\F(t) \neq \delta\,\big\}\mbox{.}
\]
Here we (implicitly) assume that all functions from $\F$ 
are finite on $t$, which means that  $\F(r)-\F(t)$ is defined.
Note that the $dd_a$ functions can be viewed as $dd_{(a,\F,\delta)}$
where $\F$ and $\delta$ are the empty tuples (i.e., $0$-tuples).

It is easy to show that all DD-functions are
\emph{bisimulation invariant}, i.e., $s \sim t$ implies
$d(s) = d(t)$ for all DD-functions $d$. So, equality of the values
of all DD-functions is a necessary condition for two states
being bisimilar. For image-finite transition systems,
this condition is also sufficient.

Let $\Delta$ be a BPP system. A key observation in~\cite{Jancar:BPP-PSPACE} 
reveals that DD-functions on states of $\Delta$ coincide with \emph{``norms''} 
w.r.t.\ effectively constructible subsets of $\C(\Delta)$. 
For all $Q \subseteq \C(\Delta)$ and $\alpha \in \P(\Delta)$ we define
\[
\mynorm_Q(\alpha) =
    \min\big\{\,\dist(\alpha,\beta) \,\mid\, \beta
    \textrm{ does not contain any constant from } Q \,\big\}\mbox{.}
\]
The result of \cite{Jancar:BPP-PSPACE} says that for every DD-function 
$d$ there is some $Q \subseteq \C(\Delta)$ such that 
$d(\alpha) =  \mynorm_Q(\alpha)$ for every $\alpha \in
\P(\Delta)$.
Since there are only finitely many subsets of $\C(\Delta)$,
there are only finitely many DD-functions which are pairwise
different on the states of $\Delta$.

So, to find out if $\alpha \sim \beta$, it suffices to
construct the relevant $Q$'s and check whether 
$\mynorm_Q(\alpha) = \mynorm_Q(\beta)$ for each of them.
Although there can be exponentially many relevant $Q$'s,
there is an algorithm performing the mentioned
checking in polynomial space
\cite{Jancar:BPP-PSPACE}.

DD-functions were also used in~\cite{JKM:BPA-BPP-bisim} 
to demonstrate the decidability of $\BPA \sim \BPP$. 
A key point was to prove that DD-functions are
\emph{prefix-encoded} over BPA processes, which, roughly speaking, means that
large finite values of DD-functions on BPA processes 
are tightly related to (i.e., represented by) large prefixes of these
processes. More precisely, given a BPA system $\Delta$,
for each DD-function $d$ there is a constant $c$ such that
if $c<d(X\alpha)<\omega$ and $X\tran{}\gamma$ then
$d(\gamma\alpha)-d(X\alpha)= \|\gamma\|-\|X\|$ (where $\|.\|$
denotes the norm, i.e.,  $\|\beta\|=\dist(\beta,\varepsilon)$).
Hence, a BPA process cannot perform a (short) sequence of moves 
causing a different change of two large finite DD-values.
We say that DD-functions are \emph{dependent} over BPA
processes,
i.e., for every two DD-functions $d_1, d_2$ there is $c$ such
that if $c<d_1(\alpha)<\omega$, $c<d_2(\alpha)<\omega$ and
$\alpha\tran{}\beta$ then 
$d_1(\beta)-d_1(\alpha)=d_2(\beta)-d_2(\alpha)$.

If we are to find out whether $\alpha \sim \beta$ for a BPA process
$\alpha$ and a BPP process $\beta$, we can proceed as follows.  By
using the above mentioned results from~\cite{Jancar:BPP-PSPACE}, one
can use standard methods from Petri net theory to show that we can
effectively check whether there are two DD-functions which are not
dependent over the states reachable from $\beta$.  If there are two
such (independent) DD-functions then $\beta$ is not bisimilar to any
BPA process. If all DD-functions are (pairwise) dependent then we can
show that there is a constant $C$ such that for every $\gamma$
reachable from $\beta$ all finite DD-values which are larger than $C$
coincide (i.e., if $c<d_1(\gamma)<\omega$ and $c<d_2(\gamma)<\omega$, then
$d_1(\gamma)=d_2(\gamma)$).
Hence, all ``large'' DD-values can be represented 
by a single number. One can even effectively construct a one-counter
process $\beta'$ which is bisimilar to $\beta$---the counter is used
to represent the ``large'' DD-values, while ``small'' DD-values are remembered
in the finite control unit. The process $\beta'$ is generally not
definable in the OC-A syntax, because there can be a need to reset the
counter back to zero in a single transition (when the ``large'' DD-values 
change to $\omega$).  However, the reset can be easily modeled in
PDA syntax by pushing a new bottom-of-stack symbol. Hence,
$\beta'$ can be seen as an (effectively definable) PDA process. 
In \cite{JKM:BPA-BPP-bisim}, the decidability proof was finished by
resorting to the involved result by S{\'{e}}nizergues 
\cite{Senizergues:PDA-bisimilarity} enabling to
verify if $\alpha\sim\beta'$. (This ``heavy machinery'' is certainly 
not necessary for establishing the decidability of
$\BPA \sim \BPP$; the reduction was used %in \cite{JKM:BPA-BPP-bisim}
just for technical convenience.)

\subsection{Undecidability Results and Lower Complexity Bounds}
\label{sec-undec-lower}

Almost all existing undecidability and hardness proofs
for simulation- and bisimilarity-checking take advantage of the
defender's ability to (indirectly) \emph{force} the attacker to do 
a specific transition. 
In a simulation game, the defender can ``threaten'' the attacker 
by a possibility to go to a universal state in the way 
indicated in Fig.~\ref{fig-sim-force} (see 
Section~\ref{sec-bisim-to-sim} for further comments). 
A similar principle can be used also in bisimulation games. 
Here, the ``threat'' is based on a possibility to enter a 
\emph{bisimilar} state. Consider processes $s,t$ with transitions 
$s \tran{a} s'$, $t \tran{a} t'$, and $t\tran{a} t''$ where $s' \sim t'$.
Under these assumptions, the move $t \tran{a} t''$ can be seen
as the only (hopeful) option available to the attacker;
the other options clearly lead to the defender's winning.
This simple idea was used implicitly, e.g., 
in~\cite{Jancar:PN-weak-high-undec}. An explicit
formulation is due to Srba \cite{Srba:BPA-weak-lower-bound-MSCS}
%PJ v tistene verzi mam v zavorce jen (Srba ), tak nevim
%v cem je problem
who used this technique to establish \PSPACE-hardness of the 
$\BPP \sim \BPP$ and $\BPA \sim \BPA$ problems 
\cite{Srba:BPP-PSPACE,Srba:BPA-PSPACE}.

To demonstrate the use (and power) of the above principles, we present 
selected undecidability and hardness proofs for concrete models. 
In Section~\ref{sec-undec} we show that the problem $\PN \wsim \PN$ is
\emph{highly} undecidable (more concretely, $\Sigma^1_1$-complete), 
and that
the problem $\PA \sm \FS$ is undecidable.

\subsubsection{Encodings of Minsky Machines.}
\label{sec-undec}

As can be expected, the undecidability results in the surveyed area
have been obtained by reductions from the halting problem.  As an
example, we will recall the result for bisimilarity over Petri nets
from~\cite{Jancar:PN-bisimilarity-TCS}. This example is not really
recent but we will expand it to show how the \emph{high}
undecidability result for \emph{weak} bisimilarity
from~\cite{Jancar:PN-weak-high-undec} can be strengthened and made much more
elegant using a recent technique of Srba ~\cite{Srba:weak-undec-compl-ENTCS}.
%PJ asi mozna aktualizace

\emph{Minsky counter machines} (with their
halting problem) are a universal model which is
technically convenient for our reduction.
\emph{A counter machine} $\M$ with nonnegative counters
$c_1,\cdots,c_m$ is a sequence of instructions
\[
  1:~\text{INS}_1;\quad 2:~\text{INS}_2;\quad \cdots \quad 
  n{-}1:\text{INS}_{n-1}; \quad n:~\texttt{halt}
\]
where each INS$_i$ ($i=1,2,...,n-1$) is in one of the following two forms
(assuming $1\leq k,l \leq n$, $1\leq j\leq m$)
\begin{itemize}
\item $c_j:=c_j+1; \texttt{ goto } k$
\item $\texttt{if } c_j = 0 \texttt{ then goto } k \texttt{ else }
              (c_j:=c_j-1; \texttt{ goto } l)$
\end{itemize}

\begin{example}
\label{exa-pn-bis-undec}
$\PN\sim\PN$ is undecidable.
\end{example}

\begin{proof}
Given a counter machine $\M$ with $m$ counters and $n$ instructions, we
construct a Petri net $\N_\M$ with places
$C_1,\dots,C_m,Q_1,\dots,Q_n,Q'_1,\dots,Q'_n$.
Intuitively, $C_1,\dots,C_m$ correspond to the counters (the
number of tokens in $C_j$ represents the value of $c_j$) and
$Q_1,\dots,Q_n$ correspond to the control places (i.e., to the
instructions)---the presence of the ``control token'' in $Q_i$ means
that INS$_i$ is now to be performed.
The places $Q'_1,\dots,Q'_n$ are ``copies'' of the control places
$Q_1,\dots,Q_n$; their purpose becomes clear later.
The (labelled) transitions of $\N_\M$ are constructed as follows.  
\begin{itemize}
\item For each instruction $i: c_j:=c_j+1; \texttt{ goto } k$ we
  add a transition depicted in
  Fig.~\ref{fig:pj-undec}~(left); an analogous transition will be also added
  for the ``copy'' places $Q'_i,Q'_k$.
\item For each instruction
 $i: \texttt{if } c_j = 0 \texttt{ then goto } k \texttt{ else }
              (c_j:=c_j-1; \texttt{ goto } l)$
  we add a transition depicted in
  Fig.~\ref{fig:pj-undec}~(middle), together with an analogous transition
  for $Q'_i,Q'_{\ell}$. We also add four transitions with label
  \textit{zer} as depicted in Fig.~\ref{fig:pj-undec}~(right).
  Note that the two ``middle'' \textit{zer}-transitions can be performed
  only when $C_j$ is positive but leave $C_j$ unchanged.
\item Finally, we add a transition
  \begin{center}
    {\small
     \psfrag{Qn}[r][r]{\mbox{$Q_n$}}
     \psfrag{hlt}[c][c]{\scriptsize\textit{hlt}}
     \includegraphics{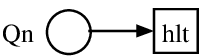}
    }
  \end{center}
  which has no counterpart for $Q'_n$.
\end{itemize}
\begin{figure}
{\small
 \psfrag{inc}[c][c]{\scriptsize\textit{inc}}
 \psfrag{dec}[c][c]{\scriptsize\textit{dec}}
 \psfrag{zer}[c][c]{\scriptsize\textit{zer}}
 \psfrag{cj}[l][l]{\mbox{$C_j$}}
 \psfrag{qi}[l][l]{\mbox{$Q_i$}}
 \psfrag{ql}[l][l]{\mbox{$Q_\ell$}}
 \psfrag{qk}[l][l]{\mbox{$Q_k$}}
 \psfrag{Qi}[r][r]{\mbox{$Q_i$}}
 \psfrag{Qk}[r][r]{\mbox{$Q_k$}}
 \psfrag{qip}[r][r]{\mbox{$Q'_i$}}
 \psfrag{qkp}[r][r]{\mbox{$Q'_k$}}
 \includegraphics{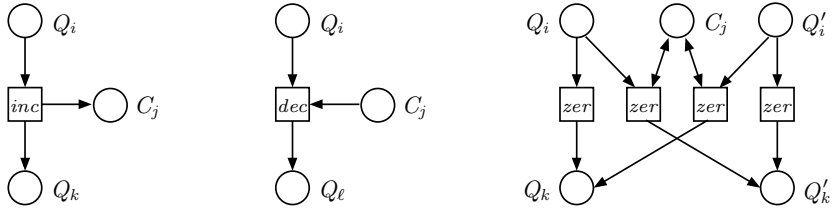}
}
\caption{Transitions of the Petri net $\N_\M$ of 
Example~\ref{exa-pn-bis-undec}}
\label{fig:pj-undec}
\end{figure}    
Having the constructed net $\N_\M$, it is a simple exercise 
to verify that the marking
with one token in $Q_1$ and zero elsewhere is bisimilar to the
marking with one token in $Q'_1$ and zero elsewhere iff 
the counter machine $\M$ halts for the zero initial values in
the counters (which is an undecidable problem).
In particular, observe the role of the previously mentioned 
forcing---if the attacker performs a move which does not correspond
to a faithful simulation of $\M$ (i.e., uses a \textit{zer}-transition 
when the respective $c_j$ is nonzero), the defender can ``punish'' him by
reaching an identical pair of markings (which is clearly a
winning position for the defender). So, the only reasonable option for
the attacker is to simulate the computation of the counter machine.
The defender must mimic, and thus the attacker wins exactly when the 
machine halts.
\end{proof}

The ``level of undecidability'' of $\PN \sim \PN$ is low; 
this is just a $\Pi^0_1$-complete problem in the arithmetical hierarchy
(the negative subcase, i.e., the existence of a winning strategy for
the attacker, is easily seen to be semidecidable).
Perhaps somewhat surprisingly, the problem  $\PN \wsim \PN$ 
turns out to be highly undecidable. In
\cite{Jancar:PN-weak-high-undec}, it was shown that the 
problem is beyond the arithmetical hierarchy, though clearly in the 
class $\Sigma^1_1$ of the analytical hierarchy.
Now we show that $\PN \wsim \PN$ is in fact a $\Sigma^1_1$-complete
problem. This is achieved by modifying the construction recently presented
by Srba \cite{Srba:weak-undec-compl-ENTCS}.

A well-known $\Sigma^1_1$-complete problem is the question whether
a given \emph{nondeterministic} counter machine allows an infinite
computation performing the first instruction infinitely often
(the ``recurrence problem''). Now we formulate another $\Sigma^1_1$-complete 
problem which better suits our purposes.

Consider ``extended'' Minsky machines which are defined in the
same way as ``ordinary'' (deterministic) Minsky machines, but the instruction
set is extended by allowing instructions of the form
\[
  i: \texttt{set } c_j\,; \texttt{ goto } k
\]
The instruction $\texttt{set } c_j$ sets the counter $c_j$ to 
a nondeterministically chosen 
value (which can be an arbitrary nonnegative integer).
Hence, we have unbounded nondeterminism. 
It is a routine programming exercise to show
that the recurrence problem can be reduced to the problem
if there is an infinite computation of our extended counter
machine: The (bounded) nondeterminism can be easily simulated;
and we can  add a special counter \textit{step} which is (programmed
to be) set to an arbitrary value before each performing of
the (original) first instruction, and 
is decremented before each other (original) instruction---if
this is not possible (since \textit{step} is 0), a jump to the halting
state is performed.

\begin{example}
\label{exa-PN-weak-high}
  $\PN\approx\PN$ is $\Sigma^1_1$-complete.
\end{example}

\begin{proof}
Let $\M$ be an extended Minsky machine. We construct a Petri net
$\N_\M$ by taking the same sets of places and transitions as
in Example~\ref{exa-pn-bis-undec}, and adding further
auxiliary places and transitions to handle instructions
of the form $i: \texttt{set } c_j\,; \texttt{ goto } k$.
The places ($r^i_1,r^i_2,r^i_3,r^i_4,r^i_5$) 
and transitions which are added for a given
instruction $i: \texttt{set } c_j\,; \texttt{ goto } k$ are
shown in Fig.~\ref{fig:pj-highundec} (their role is
explained in the following paragraphs).

\begin{figure}
{\small
 \psfrag{t}[c][c]{\mbox{$\tau$}}
 \psfrag{a}[c][c]{\mbox{$a$}}
 \psfrag{C}[r][r]{\mbox{$C_j$}}
 \psfrag{Qi}[r][r]{\mbox{$Q_i$}}
 \psfrag{Qip}[r][r]{\mbox{$Q'_i$}}
 \psfrag{Qk}[l][l]{\mbox{$Q_k$}}
 \psfrag{Qkp}[l][l]{\mbox{$Q'_k$}}
 \psfrag{ver}[c][c]{\scriptsize\textit{ver}}
 \psfrag{ri1}[l][l]{\mbox{$r^i_1$}}
 \psfrag{ri2}[l][l]{\mbox{$r^i_2$}}
 \psfrag{ri3}[l][l]{\mbox{$r^i_3$}}
 \psfrag{ri4}[l][l]{\mbox{$r^i_4$}}
 \psfrag{ri5}[l][l]{\mbox{$r^i_5$}}

 \includegraphics{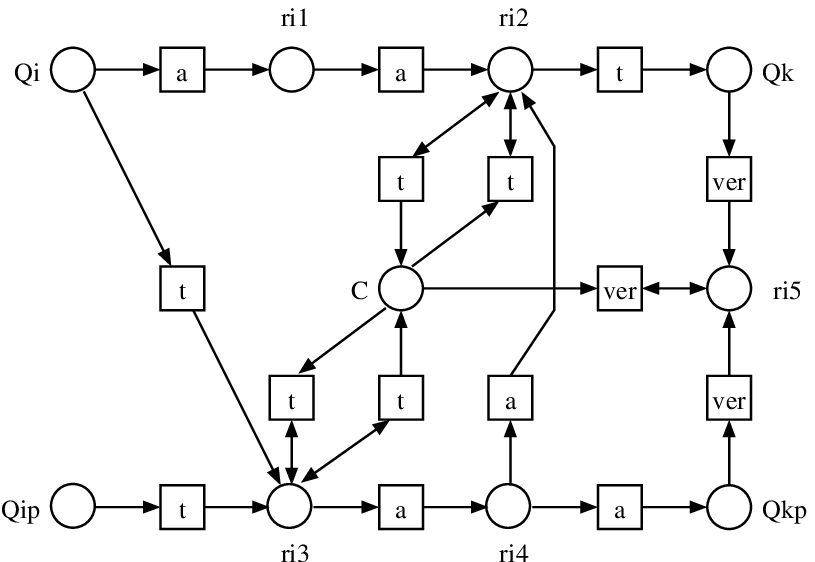}
}
\caption{Modelling the instruction 
$i: \texttt{set } c_j\,; \texttt{ goto } k$ in Example~\ref{exa-PN-weak-high}}
\label{fig:pj-highundec}
\end{figure}

Let us take two copies $\N,\N'$ of the constructed net $\N_\M$, and
assume that the control token is in $Q_i$ in $\N$ and in $Q'_i$ in
$\N'$, and the values of counters are the same in both nets.  If the
attacker wants to avoid reaching an identical pair of markings, he is
forced to start by the $a$-move from $Q_i$ in $\N$ (he moves the
control token to $r^i_1$). The defender then has to move the control
token in $\N'$ from $Q'_i$ to $r^i_4$, via the place $r^i_3$. Observe
that while having the control token in $r^i_3$, the defender could 
perform a sequence of
the respective two $\tau$-transitions and thus set any chosen value to
$C_j$ (in $\N'$).  Now, when the control tokens are in $r^i_1$ (in
$\N$) and in $r^i_4$ (in $\N'$), the attacker is forced to make the
$a$-move in $\N'$, shifting the token from $r^i_4$ to $Q'_k$
(otherwise the defender could immediately reach an identical pair of
markings). The defender answers by moving the token from $r^i_1$
to $Q_k$ (in $\N$) via $r^i_2$, where he can set $C_j$ (in $\N$) to
any chosen value. (We can safely assume that the instruction $k$ is not
another set-instruction and thus no $\tau$-moves are possible from
$Q_k$, $Q'_k$. The defender does not gain anything by leaving the token in
$r^i_2$, because the attacker could move the token to $Q_k$ in the next 
round anyway.)  Now, the control tokens are in $Q_k$, $Q'_k$ and it was
the defender who set values to $C_j$ in both $\N$, $\N'$. If the defender 
has set two different values, the attacker can obviously win by performing
a sequence of actions \textit{ver}. Otherwise, the correct simulation
of a computation of $\M$ continues.

Hence, starting with markings $M$ of $\N$ and $M'$ of $\N'$, 
where $M$ and $M'$ has just a token in $Q_1$ and $Q'_1$, respectively,
it is clear that $M \wsim M'$ iff $\M$ has an infinite computation.
\end{proof}

Reductions of the halting problem to simulation problems are usually
simpler, because the constructed processes do not have to be 
``coupled'' so tightly as in the case of bisimilarity. This is demonstrated
in the last example of this subsection.

\begin{example}
 $\PA \sm \FS$ is undecidable.
\end{example}

\begin{proof}
Let $\M$ be a counter machine with two 
counters initialized to zero. We construct a (deterministic)
PA process $Z_1 \| Z_2$ and a deterministic FS process
$f_1$ such that $Z_1 \| Z_2 \sm f_1$ iff $\M$ does not halt.

The rules of the underlying system of $Z_1\|Z_2$ look as follows:
\[
\begin{array}{llll}
Z_1 \tran{z_1} Z_1, & 
Z_1 \tran{i_1} C_1.Z_1, & 
C_1 \tran{i_1} C_1.C_1, & 
C_1 \tran{d_1} \varepsilon, \\
Z_2 \tran{z_2} Z_2, &
Z_2 \tran{i_2} C_2.Z_2, &
C_2 \tran{i_2} C_2.C_2, &
C_2 \tran{d_2} \varepsilon
\end{array}
\]
Hence, $Z_1 \| Z_2$ is a parallel composition of two counters initialized
to zero. The underlying FS system $\Delta$ of $f_1$ corresponds to the finite 
control of $\M$. For every instruction of the form
$i: c_j := c_j{+}1;~\texttt{goto }~k$ we have a rule $f_i \tran{i_j} f_k$.
For every instruction of the form
$i: \texttt{if}\ c_j = 0\ \texttt{then goto}\ k\ \texttt{else}\ 
   c_j := c_j{-}1;\ \texttt{goto}\ l$
we have the rules $f_i \tran{z_j} f_k$ and 
$f_i \tran{d_j} f_l$. Then we ``enforce'' these transitions. That is,
\begin{itemize}
\item we add a new constant $u$ together with rules 
  $u \tran{a} u$ for every action $a$;
\item for every $f_i$, where $i<n$, and every action $a$: If there is no 
  rule $f_i \tran{a} f_j$ for any $f_j$, then we add a rule $f_i \tran{a} u$. 
\end{itemize}
The attacker (who plays with $Z_1\|Z_2$) can choose a counter and perform
one of the available operations on it. Since the defender ``enforces''
the right choice, the only attacker's chance is to faithfully emulate 
the machine $\M$; if $\M$ halts, then the defender is eventually forced
to enter the state $f_n$ where he loses the game. Hence, 
$Z_1\|Z_2 \sm f_1$ iff $\M$ does not halt.  
\end{proof}

\subsubsection{Hardness Results.} 

The use of the ``enforced'' transitions in hardness proofs will be
demonstrated on two examples. We show that the problems 
$\PDA \sim \FS$ and $\PDA \sm \FS$ are \PSPACE-hard
by reducing the QBF (Quantified Boolean Formula)
problem to each of them. Our objective is to
show what has to be done differently in the two respective cases,
i.e., how the two ``enforcing'' techniques are implemented
for the same models.
(Note that the problems $\PDA \sim \FS$ and $\PDA \sm \FS$ are 
in fact \PSPACE-complete and \EXPTIME-complete, respectively 
\cite{KM:PDA-BPA-sim-bisim}). 

For the rest of this section, let us fix a quantified Boolean formula
\[
   \varphi \ \equiv \ 
   \forall x_1 \exists x_2 \cdots \forall x_{n-1} \exists x_n : 
      C_1 \wedge \cdots \wedge C_m
\]   
where every $C_i$ is a clause, i.e., a disjunction of possibly negated 
propositions from $\{x_1,\dots,x_n\}$. We can safely assume that $n$ is even.
The problem whether a given quantified Boolean formula
holds is known to be \PSPACE-complete; see, e.g., \cite{Papadimitriou:book}.

\begin{example}
$\PDA \sm \FS$ is \PSPACE-hard.
\end{example}
\begin{proof}
  Let us consider a process $g L_1 Z$ of a PDA system with rules
  \begin{itemize}
  \item $gL_i \tran{a} gL_{i+1} X_i$, $gL_i \tran{a} gL_{i+1} \bar{X}_i$
    for all odd $i$ such that $1 \leq i < n$;
  \item $gL_i \tran{b} gL_{i+1} X_i$, $gL_i \tran{c} gL_{i+1} \bar{X}_i$
    for all even $i$ such that $1 \leq i \leq n$;
  \item $gL_{n+1} \tran{d} c_j \varepsilon$ for every $1 \leq j \leq m$;
  \item $c_j X_i \tran{d} c_j X_i$, $c_j \bar{X}_i \tran{d} c_j \varepsilon$
    for all $1 \leq i \leq n$ and $1 \leq j \leq m$ such that 
    $x_i$ appears in the clause $C_j$;   
  \item $c_j X_i \tran{d} c_j \varepsilon$, 
    $c_j \bar{X}_i \tran{d} c_j \bar{X}_i$
    for all $1 \leq i \leq n$ and $1 \leq j \leq m$ such that 
    $\neg x_i$ appears in the clause $C_j$;   
  \item $c_j Z \tran{e} c_j Z$ for all $1 \leq j \leq m$.
  \end{itemize}
  We claim that the fixed quantified Boolean 
  formula $\varphi$ holds iff $g L_1 Z \sm f$, where
  $f$ is a finite-state process of the following system:
  \begin{center}
    \psfrag{f}[c][c]{\small\mbox{$f$}}
    \psfrag{la}[l][l]{\tiny\mbox{$a$}}
    \psfrag{ra}[r][r]{\tiny\mbox{$a$}}
    \psfrag{b}[l][l]{\tiny\mbox{$b$}}
    \psfrag{c}[r][r]{\tiny\mbox{$c$}}
    \psfrag{d}[l][l]{\tiny\mbox{$d$}}
    \includegraphics{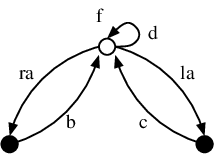} 
  \end{center}
  Here, the black-filled circles denote the states which enforce
  the actions of their outgoing transitions (see Section~\ref{sec-sim-bisim}).
  Intuitively, the attacker (who plays with $gL_1 Z$) is responsible 
  for choosing the assignment for variables with odd index, while the
  defender (who plays with $f$) chooses the assignment for variables 
  with even index by forcing the attacker to do $b$ or $c$ in the next
  round. After the guessing phase, the attacker chooses a clause
  by performing one of the $gL_{n+1} \tran{d} c_j \varepsilon$ transitions
  and starts to pop symbols from the stack, trying to find a symbol which
  witnesses the validity of the chosen clause. If no such symbol is found,
  the attacker eventually emits the action $e$ and thus wins the game. 
  Otherwise, he just performs an infinite number of $d$'s and hence the 
  defender wins.
\end{proof}

\begin{example}
  The problem $\PDA \sim \FS$ is \PSPACE-hard.
\end{example}
\begin{proof}
  For purposes of this proof, let us assume (wlog) 
  that $\varphi$ contains a clause which is true for every assignment.
  Let $g L_1 Z$ be a PDA process defined by
  \begin{itemize}
  \item $gL_i \tran{a} gL_{i+1} X_i$, $gL_i \tran{a} gL_{i+1} \bar{X}_i$
    for all $1 \leq i \leq n$;
  \item $gL_{n+1} \tran{c} c_j \varepsilon$ for every $1 \leq j \leq m$;
  \item $c_j X_i \tran{d} p \varepsilon$, 
        $c_j \bar{X}_i \tran{d} c_j \varepsilon$
    for all $1 \leq i \leq n$ and $1 \leq j \leq m$ such that 
    $x_i$ appears in the clause $C_j$;   
  \item $c_j X_i \tran{d} c_j \varepsilon$, 
    $c_j \bar{X}_i \tran{d} p \varepsilon$
    for all $1 \leq i \leq n$ and $1 \leq j \leq m$ such that 
    $\neg x_i$ appears in the clause $C_j$;
  \item $p X_i \tran{d} p \varepsilon$, $p \bar{X}_i \tran{d} p \varepsilon$
    for all $1 \leq i \leq n$;    
  \item $c_j Z \tran{e} c_j Z$ for all $1 \leq j \leq m$.
  \end{itemize}
  Moreover, we also add transitions $gL_i \tran{a} \bar{f}_{i+1} L_i$ 
  for every even $i$ where $1 \leq i \leq n$, and another family of 
  transitions which ensure
  that every process of the form $\bar{f}_{i+1} L_i \alpha$, where
  $1 \leq i \leq n$, is bisimilar
  to the state $\bar{f}_{i+1}$ in the following finite-state system:
  \begin{center}
    \psfrag{f1}[c][c]{\small\mbox{$f_1$}}
    \psfrag{f2}[c][c]{\small\mbox{$f_2$}}
    \psfrag{f3}[c][c]{\small\mbox{$f_3$}}
    \psfrag{f4}[c][c]{\small\mbox{$f_4$}}
    \psfrag{f5}[c][c]{\small\mbox{$f_5$}}
    \psfrag{fn1}[c][c]{\small\mbox{$f_{n+1}$}}
    \psfrag{bfn1}[c][c]{\small\mbox{$\bar{f}_{n+1}$}}
    \psfrag{bf1}[c][c]{\small\mbox{$\bar{f}_1$}}
    \psfrag{bf2}[c][c]{\small\mbox{$\bar{f}_2$}}
    \psfrag{bf3}[c][c]{\small\mbox{$\bar{f}_3$}}
    \psfrag{bf4}[c][c]{\small\mbox{$\bar{f}_4$}}
    \psfrag{bf5}[c][c]{\small\mbox{$\bar{f}_5$}}
    \psfrag{sc}[c][c]{\small\mbox{$c$}}
    \psfrag{bsc}[c][c]{\small\mbox{$\bar{c}$}}
    \psfrag{g1}[c][c]{\small\mbox{$g_1$}}
    \psfrag{bg1}[c][c]{\small\mbox{$\bar{g}_1$}}
    \psfrag{g2}[c][c]{\small\mbox{$g_2$}}
    \psfrag{bg2}[c][c]{\small\mbox{$\bar{g}_2$}}
    \psfrag{gn1}[c][c]{\small\mbox{$g_{n+1}$}}
    \psfrag{bgn1}[c][c]{\small\mbox{$\bar{g}_{n+1}$}}
    \psfrag{a}[c][c]{\tiny\mbox{$a$}}
    \psfrag{c}[c][c]{\tiny\mbox{$c$}}
    \psfrag{d}[c][c]{\tiny\mbox{$d$}}
    \psfrag{e}[c][c]{\tiny\mbox{$e$}}
    \psfrag{ra}[r][r]{\tiny\mbox{$a$}}
    \psfrag{lc}[l][l]{\tiny\mbox{$c$}}
    \includegraphics{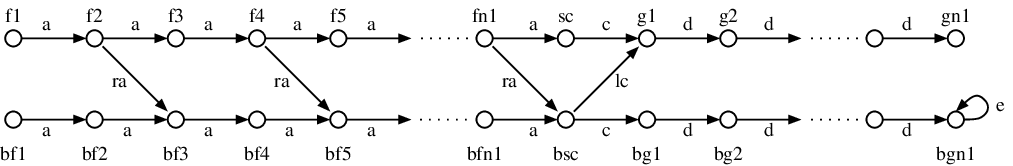} 
  \end{center}
  We argue that $\varphi$ holds iff $g L_1 Z \sim f_1$.  
  The ``ideal'' scenario for bisimulation game between the two processes
  looks as follows: the 
  assignment for variables with
  odd index is chosen by the attacker who performs an appropriate $a$-move 
  in the PDA process; the defender  has to reply by the only available 
  $a$-move in the finite-state system. If a variable $x_i$ with an even 
  index is to be assigned a value, the attacker performs the move 
  $f_i \tran{a} f_{i+1}$ in the finite-state system. Now we distinguish 
  two possibilities.
  \begin{itemize}
    \item the formula 
       $\exists x_i \forall x_{i+1} \cdots \exists x_n : 
       C_1 \wedge \cdots \wedge C_m$
       is false after substituting each occurrence of $x_j$ (for all $j < i$) 
       with its previously assigned value. Then, the defender chooses
       some assignment for $x_i$ by performing an $a$-move in the PDA process, 
       but it does not really matter which 
       one---from this point on, the attacker can always 
       choose such an assignment for variables with odd index so that
       the above given formula is false for every even $i$. Hence,
       the attacker can enforce the game situation when one token is on $c$ and
       the chosen assignment falsifies some clause $C_j$. Then, the
       attacker performs the transition $gL_{n+1} \tran{c} c_j \varepsilon$
       and the defender has to respond by $c \tran{c} g_1$. Now, the
       attacker pops symbols from the stack, and since there is no
       symbol witnessing the validity of $C_j$, he eventually emits $e$
       and thus he wins.
    \item otherwise, the defender chooses the ``right'' value for $x_i$,
       keeping a chance that the final assignment will satisfy
       all clauses. If the formula $\varphi$ holds, he can thus
       enforce the game situation when one token is on $c$ and the
       assignment stored in the PDA processes satisfies every clause $C_j$;
       it is easy to check that the defender wins the game from this 
       configuration.
   \end{itemize}     
  The construction ensures that the two players do not gain anything
  by violating the just specified scenario (a full justification 
  requires a detailed analysis). For example,
  the attacker cannot use the transitions $f_{i} \tran{a} \bar{f}_{i+1}$ in the
  finite-state system because the defender could go to a bisimilar PDA state. 
\end{proof}

\section{An Overview of Existing Results}
\label{sec-results}

In this section we give a brief overview of existing decidability
and complexity results from the area of
equivalence-checking on infinite-state
processes. Results about the related \emph{regularity} problem are
also presented (given a process $s$ and a behavioral equivalence
$\leftrightarrow$, we ask if $s$ is ``regular'', i.e., equivalent
to some unspecified finite-state process).

\begin{figure}
{\scriptsize
\psfrag{FS}[c][c]{$\FS$}
\psfrag{BPP}[c][c]{$\BPP$}
\psfrag{BPA}[c][c]{$\BPA$}
\psfrag{OCN}[c][c]{$\OCN$}
\psfrag{OCA}[c][c]{$\OCA$}
\psfrag{PA}[c][c]{$\PA$}
\psfrag{PN}[c][c]{$\PN$}
\psfrag{PDA}[c][c]{$\PDA$}
\psfrag{PPDA}[c][c]{$\PPDA$}
\psfrag{eq1}[l][l]{$\sim \FS$}
\psfrag{eq2MMMMMMMMMMMMMMMMMMMMMMMMM}[l][l]{$\sm \FS$, $\FS \sm$, $\sme \FS$}
\psfrag{eq3}[l][l]{$\wsim \FS$}
\psfrag{eq4}[l][l]{$\sim$}
\psfrag{eq5}[l][l]{$\sm$, $\sme$}
\psfrag{eq6}[l][l]{$\wsim$}
\hspace*{2ex}\includegraphics{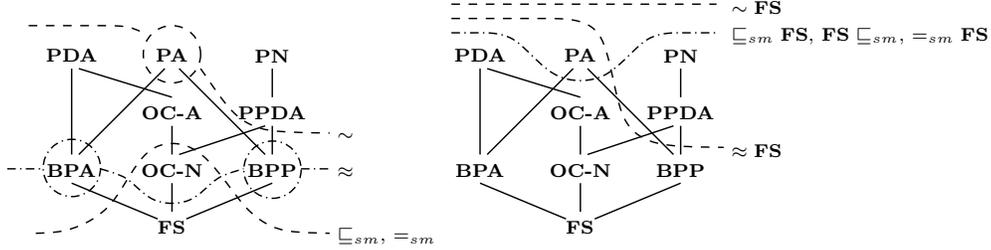}
}
\caption{The Decidability Border for Equivalence-Checking Problems}
\label{fig-border}
\end{figure}

The decidability border for equivalence-checking on infinite-state
processes has already been determined for some behavioral
equivalences. The left-hand part of Fig.~\ref{fig-border} shows the
decidability border for the problem $\textbf{C} \leftrightarrow
\textbf{C}$, where $\textbf{C}$ is a subclass of PRS and
$\leftrightarrow$ one of the $\sim$, $\wsim$, and $\sme$ equivalences
(the decidability of $\PA \sim \PA$, $\BPA \wsim \BPA$, and $\BPP
\wsim \BPP$ is still open; this is indicated by dashed circles because
it is not known whether the bordering line goes above or below the
considered class).  The right-hand side of Fig.~\ref{fig-border} shows
the decidability border for the $\textbf{C} \leftrightarrow
\textbf{FS}$ problem.  Detailed comments are split into several
subsections.

\subsection{Results for (Weak) Bisimilarity}

\subsubsection{Bisimilarity-Checking between Infinite-State Systems}
The first result indicating that bisimilarity is ``more decidable'' than
trace/language equivalence is due to Baeten, Bergstra, and Klop
\cite{BBK:nBPA-bisimilarity-JACM} who established the decidability of
bisimilarity for normed BPA processes. The proof is based on isolating
a complex periodicity hidden in the structure of transition systems
generated by normed BPA processes.  A simpler proof of this result was
later given by Caucal in \cite{Caucal:selfbisimulation-RAIRO}, where
the technique of bisimulation bases was introduced. Another short proof
is \cite{Groot:BPA-bisimilarity}.  In \cite{HS:nBPA-bisimilarity-JLC},
a sound and complete tableau-based deductive system for bisimilarity
on normed BPA processes has been designed. The complexity of the
problem was first addressed by Huynh and Tian
\cite{HT:nBPA-bisimilarity-TCS} who gave a $\Sigma_2^P =
\NP^{\mathbf{NP}}$ upper bound. Later, Hirshfeld, Jerrum, and Moller
demonstrated that the problem is decidable in polynomial time
\cite{HJM:BPA-polynomial-TCS}.  The decidability result has been
extended to all (not necessarily normed) BPA processes by Christensen,
H\"{u}ttel, and Stirling in \cite{CHHS:BPA-bisimilarity-IC}. Again, it
is shown that bisimilarity over all states of a given BPA system
can be represented by a finite bisimulation base. As the
decidability
result is obtained by a combination of two semidecision procedures, it
does not allow for any complexity estimations. An algorithm with
elementary complexity was given in \cite{BCS:BPA-bisim-elementary}
(the authors mention that some straightforward optimizations would
lead to a doubly exponential algorithm). A technical core of the
result is a procedure which computes a finite bisimulation base for
general BPA processes.  Recently, a \PSPACE\ lower bound for the problem
$\BPA \sim \BPA$ has been established by Srba in
\cite{Srba:BPA-PSPACE}. The exact complexity classification is still missing.

The observation that bisimilarity over processes of a given BPP system
is finitely generated by a bisimulation base is due to 
Christensen, Hirshfeld, and Moller \cite{CHHM:BPP-bisimilarity}
who proved the decidability of bisimilarity for BPP processes.
A polynomial-time algorithm for normed BPP processes has been given
in \cite{HJM:nBPP-polynomial-MSCS}. The complexity of the general
case was addressed by Mayr in \cite{Mayr:BPP-bis-hardness} who gave
a \coNP-lower bound for the problem, which has been improved
to \PSPACE\ by Srba \cite{Srba:BPP-PSPACE}. This result has 
recently been complemented by Jan\v{c}ar who gave a matching \PSPACE\ upper 
complexity bound \cite{Jancar:BPP-PSPACE}, which means that
the $\BPP \sim \BPP$ problem is \PSPACE-complete. When Jan\v{c}ar's 
algorithm is carefully implemented for normed BPP processes, it runs in 
time $O(n^3)$, as shown in \cite{JK:nBPP-new-PTIME}. 

The decidability of bisimilarity between normed BPA and normed BPP processes
was proved by Blanco \cite{Blanco:normed} and independently
in \cite{CKK:BPAvsBPP-AI}. Later, the result was extended to 
parallel compositions of normed BPA and normed BPP processes in
\cite{Kucera:nBPA-parallelize-TCS}. Recently, the decidability
of $\BPA \sim \BPP$ has been established in \cite{JKM:BPA-BPP-bisim}.
A deep result \cite{HJ:nPA-bisimilarity}
due to Hirshfeld and Jerrum says that bisimilarity is decidable for normed
PA processes. The proof is based on the unique decomposition property
of normed processes w.r.t.{} ``$.$'' and ``$\|$'', and hence the
method is not applicable to general PA processes. 

The semilinear structure of bisimilarity over one-counter processes
has been identified in \cite{Jancar:one-counter-IC}; it allows to
conclude that bisimilarity is semidecidable (and thus decidable) for
one-counter processes. However, the problem is computationally intractable
even for one-counter nets---\DP-hardness of $\OCN \sim \OCN$ was demonstrated
in \cite{Kucera:OC-FS-weak-bisimilarity-TCS} (the class \DP\ is expected
to be somewhat larger than the union of $\NP$ and $\coNP$). In 
\cite{Senizergues:PDA-bisimilarity}, S{\'{e}}nizergues proved that
bisimilarity is decidable for general PDA processes. 
This also extends a previous result
due to Stirling \cite{Stirling:nPDA-bisimilarity-TCS} which says that
bisimilarity is decidable for a subclass of PDA processes which can always
empty their stack. 
S{\'{e}}nizergues's proof is obtained by adapting the method which
previously led to the decidability of language equivalence for
deterministic pushdown automata \cite{Senizergues:LALB-TCS}.
Recently, Stirling presented a primitive recursive algorithm for the same
problem \cite{Stirling:DPDA-language-prip-rec}. As for lower 
bounds, the $\PDA \sim \PDA$ problem is known to be \EXPTIME-hard
\cite{KM:PDA-BPA-sim-bisim}.

The undecidability of bisimilarity for Petri nets is due to 
Jan\v{c}ar \cite{Jancar:PN-bisimilarity-TCS}. In fact, the proof 
(see Example~\ref{exa-pn-bis-undec}) also works for PPDA processes.
A related undecidability result is \cite{Schnoebelen:LCS} where 
Schnoebelen proved that bisimilarity as well as other process equivalences
are undecidable for lossy channel systems. 

As for weak bisimilarity, many problems are still open. Weak
bisimilarity is known to be semilinear, and thus semidecidable for BPP
processes \cite{Esparza:weak-semidec}. 
Although the general case is
still open, there is a decidability result for the subclass of
\emph{totally normed} BPP processes
\cite{Hirshfeld:tnBPA-tnBPP-weak-ENTCS} (a process is totally normed
if it can reach $\varepsilon$ in a finite sequence of transitions, but
each such sequence must contain at least one action different from
$\tau$). The best known lower bound for the $\BPP \wsim
\BPP$ problem is \PSPACE\ \cite{Srba:BPA-weak-lower-bound-MSCS}, which is
valid also for the normed subcase (previously, there was an \NP\ 
\cite{Stribrna:BPA-BPP-weak-hardness-ENTCS} and $\Pi_2^P =
\coNP^{\textbf{NP}}$ lower bound \cite{Mayr:BPP-bis-hardness}). Weak
bisimilarity between totally normed BPA processes is also decidable
\cite{Hirshfeld:tnBPA-tnBPP-weak-ENTCS}. The problem
$\BPA \wsim \BPA$ is known to be \PSPACE-hard
\cite{Stribrna:BPA-BPP-weak-hardness-ENTCS}, even in the normed
subcase \cite{Srba:BPA-weak-lower-bound-MSCS}.
Recently, the lower complexity
bound for weak bisimilarity on normed BPA has been improved to $\EXPTIME$ in 
\cite{Mayr:BPA-EXPTIME-ENTCS}.
The problem $\PDA \wsim \PDA$ is already undecidable 
\cite{Srba:PDA-weak-undecidable}. This result has been 
generalized in \cite{Mayr:OC-weak-undec} where it is shown that 
even the problem $\OCN \wsim \OCN$ is undecidable. An incomparable result
of \cite{Srba:PA-weak-undec} shows that
$\PA \wsim \PA$ is also undecidable \cite{Srba:PA-weak-undec}.
Weak bisimilarity between Petri nets is even \emph{highly}
undecidable (i.e., beyond arithmetical hierarchy)
\cite{Jancar:PN-weak-high-undec}; this result has been
strengthened to $\Sigma^1_1$-completeness and achieved
also for PDA and PA in \cite{JS:high-undec-PA}.

\subsubsection{Bisimilarity-Checking between an Infinite and a Finite-State 
System}

The problem has been
considered in \cite{JM:PN-properties} where it is shown that
$\PN \sim \FS$ is decidable. However, $\PN \wsim \FS$ is already
undecidable \cite{EJ:PN-regularity}.
The decidability of $\BPP \wsim \FS$ was shown in \cite{Mayr:BPP-weak}.
Theorem~\ref{thm-bis-FS} has been explicitly formulated in 
\cite{JK:FS-bisim-ENTCS} and (in a more abstract form) in 
\cite{JKM:bisim-like-TCS} where it is also shown that weak bisimilarity
is decidable between so-called PAD processes and finite-state
ones (the PAD class subsumes both PA and PDA processes). Complexity
results followed---in \cite{KM:BPA-nBPP-FS-weak-TCS} it was shown
that the problems $\BPA \wsim \FS$ and $\nBPP \wsim \FS$ are solvable
in polynomial time. The problem $\BPP \wsim \FS$ is in \PSPACE\ 
\cite{JKM:bisim-like-TCS}, and the problem  $\BPP \sim \FS$ is in
\PTIME \cite{KS:BPP-FS-PTIME}. The problem $\PDA \sim \FS$ is \PSPACE-hard 
\cite{Mayr:PDA-FS-bis-hard}, and the matching upper bound for 
$\PDA \wsim \FS$ was given in \cite{KM:PDA-BPA-sim-bisim},
which means that the problems $\PDA \sim \FS$ and  $\PDA \wsim \FS$
are \PSPACE-complete. Bisimilarity between one-counter processes
and finite-state processes was studied in 
\cite{Kucera:OC-FS-weak-bisimilarity-TCS}. It is shown that $\OCN \wsim \FS$
is \DP-hard, while $\OCA \sim \FS$ is solvable in polynomial time.
The decidability of bisimilarity between lossy channel systems 
and finite-state systems is due to \cite{AK:LCSvsFS}.
However, this problem (and in fact all non-trivial problems related
to formal verification of lossy channel systems) are of nonprimitive
recursive complexity \cite{Schnoebelen:LCM-non-primitive}.

\subsubsection{Regularity-Checking}
The decidability of regularity w.r.t.{} $\sim$ for Petri nets
is due to \cite{EJ:PN-regularity}. The regularity problem is also
decidable for BPA processes \cite{BCS:BPA-regularity}
and \mbox{OC-A} processes \cite{Jancar:one-counter-IC}. 
For normed processes, regularity w.r.t.{} $\sim$ usually coincides
with ``syntactical boundedness'', i.e., the question if a given process
can reach infinitely many syntactically distinct states. This condition
can be in some cases checked in polynomial time; it applies, e.g., to
normed PA \cite{Kucera:PA-regularity-IPL} and normed PDA processes.
There are also some lower complexity bounds---regularity-checking 
w.r.t.{} $\sim$ is known
to be \PSPACE-hard for BPA \cite{Srba:BPA-PSPACE} and BPP 
\cite{Srba:BPP-PSPACE} (previously, there was  \coNP-lower bound
for BPP \cite{Mayr:BPP-bis-hardness} and \PSPACE-lower
bound for PDA \cite{Mayr:PDA-FS-bis-hard}). For Petri nets, one can easily
establish the \EXPSPACE-lower bound by employing the simulation
of a deterministic exponentially bounded machine due to Lipton
\cite{Lipton:PN-Reachability}. 
The problem is still open
for general PA and PDA processes, though it is clearly semidecidable
because bisimilarity with a (given) finite-state process is decidable
for these models. Regularity w.r.t.{} $\wsim$
is undecidable for Petri nets \cite{EJ:PN-regularity} and 
$\EXPTIME$-hard for PDA \cite{Mayr:BPA-EXPTIME-ENTCS};
for other major models of infinite-state systems, the problem remains
open (it is again at least semidecidable by applying the same 
argument as above).

\subsection{Results for Simulation and Trace Preorder/Equivalence}

\subsubsection{Simulation Preorder/Equivalence}
As opposed to bisimilarity, simulation preorder/equivalence between
infinite-state processes tends to be undecidable. Since trace preorder
and simulation preorder coincide over deterministic processes, the
undecidability of simulation preorder/equivalence for BPA processes
follows immediately from Friedman's result 
\cite{Friedman:simple-grammars-incl}
which says that the language inclusion problem for simple grammars 
is undecidable. As for BPP, simulation preorder/equivalence is also
undecidable as shown by Hirshfeld \cite{Hirshfeld:PN-equivalence}.
The only known class of infinite-state processes where simulation
preorder/equivalence remains decidable are one-counter nets. The
result has been achieved by Abdulla and  {\v{C}}er{\={a}}ns 
\cite{AC:one-counter-simulation}. A simpler proof was later given
in \cite{JMS:OC-simulation}, where it is also shown that simulation
preorder/equivalence for one-counter processes is already undecidable. 
A \DP\ lower bound for the $\OCN \sm \OCN$ and $\OCN \sme \OCN$
problems is given in \cite{JKMS:one-counter-generic-IC}.

Deciding simulation between an infinite and a finite-state system is 
computationally easier. The decidability of $\PN \sm \FS$, $\FS \sm \PN$
(and thus also $\PN \sme \FS$) is due to
\cite{JM:PN-properties}. Simulation between lossy channel systems and
finite systems is also decidable (in both directions) \cite{AK:LCSvsFS}.
The result of \cite{Schnoebelen:LCM-non-primitive} implies
that this problem is of nonprimitive recursive complexity
A more general argument showing the decidability of simulation
between processes of the so-called well-structured transition systems
and finite-state processes has been presented in \cite{ACJT:general-theorems}.

The decidability/tractability border 
for the problem has been established in \cite{KM:simulation-FS-IC}. It is 
shown that $\PDA \sm \FS$ and $\FS \sm \PDA$ are in \EXPTIME, and that
$\PA \sm \FS$ and $\FS \sm \PA$ are already undecidable. Moreover,
the following lower bounds are given:
$\FS \sm \BPA$ and $\FS \sm \BPP$ are \PSPACE-hard, and
$\BPA \sm \FS$ and $\BPP \sm \FS$ (thus also
for $\BPA \sme \FS$ and $\BPP \sme \FS$) are \coNP-hard. Recently
\cite{KM:PDA-BPA-sim-bisim}, the simulation preorder/equivalence
problem between a BPA/PDA process and a finite-state process was shown
to be \EXPTIME-complete (for both directions of simulation preorder).
In this case, the only difference between PDA and BPA 
(from the complexity point of view) is that simulation preorder/equivalence
between PDA and FS is \EXPTIME-complete even for a \emph{fixed} 
finite-state process, while simulation between a BPA and any 
fixed finite-state process $f$ is decidable in polynomial 
time \cite{KM:PDA-BPA-sim-bisim}.
Other tractable problems are $\OCN \sm \FS$, $\FS \sm \OCN$, and
$\OCN \sme \FS$, which are all decidable in polynomial time 
\cite{Kucera:simulation-seq-systems}. However, $\OCA \sm \FS$, $\FS \sm \OCA$, 
and $\OCA \sme \FS$ are already \DP-hard 
\cite{Kucera:simulation-seq-systems,JKMS:one-counter-generic-IC}. 
As for regularity-checking w.r.t.{} $\sme$, the problem
is known to be decidable for OC-N processes 
\cite{JKM:one-counter-sim-bisim}, and undecidable for Petri nets
\cite{JM:PN-properties} and PA processes \cite{KM:simulation-FS-IC}. 

\subsubsection{Trace Preorder/Equivalence} 
Since trace preorder/equivalence are closely related to language
inclusion/equivalence of automata theory \cite{HU:book}, all 
(un)decidability results about BPA and PDA processes follow easily from
the ``classical'' ones. It means that almost all problems are undecidable;
the only notable exception is the $\PDA \tr \FS$ problem which is decidable.
The undecidability of trace preorder/equivalence between BPP processes
is due to \cite{Hirshfeld:PN-equivalence}. 
%% Of course, some of the generally undecidable problems become decidable
%% when restricted to deterministic processes; for example, trace
%% equivalence for general BPP is undecidable, but it is
%% decidable for deterministic BPP processes because it
%% coincides with bisimilarity over the class of deterministic~processes.

Trace preorder/equivalence with a finite-state system is undecidable
for BPA and PDA, but decidable for Petri nets; $\PN \tr \FS$ and
$\FS \tr \PN$ are decidable as shown in \cite{JM:PN-properties}. In the
same paper it is shown that regularity w.r.t.{} $\tre$
is undecidable for~Petri~nets.  

\subsubsection{Acknowledgment} 
We thank Ji\v{r}\'{\i} Srba for his many useful comments and suggestions.

\bibliographystyle{acmtrans}
\bibliography{str-long,concur}

\end{document}